# Real-time reflectance generation for UAV multispectral imagery using an onboard downwelling spectrometer in varied weather conditions

Jiayang Xie[a,b,1], Yutao Shen[a,b,1], Haiyan Cen[a,b,*]

[a] College of Biosystems Engineering and Food Science, Zhejiang University, Hangzhou 310058, China

[b] Key Laboratory of Spectroscopy Sensing, Ministry of Agriculture and Rural Affairs, Hangzhou 310058, China

* Corresponding author at: College of Biosystems Engineering and Food Science, Zhejiang University, Hangzhou 310058, China.

   E-mail address: hycen@zju.edu.cn (H. Cen).

[1] Co-first authors.

## Abstract

   Advancements in unmanned aerial vehicle (UAV) remote sensing with spectral imaging enable efficient assessment of critical agronomic traits. However, existing reflectance calibration or generation methods often suffer from limited prediction accuracy and practical flexibility. This study investigates the integration of an on-board downwelling spectrometer (DS) with a multispectral imager (MSI) to convert digital number (DN) values into reflectance using real-time solar spectra as references. To ensure consistent measurements of incident light, an upward gimbal-mounted DS was attached to the UAV, and a sinusoidal model was developed to correct for solar position variability. Reflectance was predicted using principal component regression (PCR) with the four major principal components (PCs) extracted from the reference solar spectrum (Top-4PC PCR model). This model performed equivalently to the direct correction method, surpassing the empirical line method (ELM) by 86.1% in root mean square error (RMSE) under fluctuating cloudy conditions and outperforming the downwelling light sensor (DLS) method by 59.0% across varying illumination conditions. Additionally, a novel 4-Band multiple linear regression (MLR) model, utilizing only four sensitive bands with a 30 nm bandwidth, achieved the same performance as the Top-4PC PCR model and the direct correction method. In two diurnal validations, the RMSE was 2.24% in a ground experiments under partial cloud cover, and 2.03% in a UAV campaign conducted at various times throughout a sunny day. Furthermore, this model proved effective in a UAV imaging campaign over a large rice field with substantial cloud fluctuations, achieving a 95.0% improvement in canopy reflectance consistency within a homogeneous vegetation area. Evaluation of vegetation indices—normalized difference vegetation index (NDVI) and difference vegetation index (DVI), representing ratio and non-ratio indices respectively—showed improvements in consistency by 86.0% and 90.3%. The validations indicate that this cost-effective method with lower hardware requirements exhibits robust performance in extending UAV spectral sensing operation windows and standardizing data collection across diverse weather conditions.

# 1. Introduction

Unmanned aerial vehicle (UAV) remote sensing offers a cost-effective method for collecting diverse data from various sensors, providing valuable insights for agriculture, environmental monitoring, and more (Zhang and Zhu, 2023). Advancements in UAV technologies offer valuable opportunities to assess critical factors influencing crop yield and quality. It operates as a high-throughput, nondestructive method for evaluating a wide range of plant phenotypes, including morphological traits (e.g., plant height and leaf area index), biochemical traits (e.g., chlorophyll and nitrogen contents), physiological traits (e.g., photosynthesis and stress resistance), and performance traits (e.g., biomass and yield) (Sun et al., 2022). Compared to handheld or ground vehicle-based measurements, UAV technologies provides more efficient deployment and higher throughput, and achieves higher spatial-temporal resolution than satellite imagery, thereby serving as a valuable tool for exploring the interaction between genotype and environment and advancing crop breeding efforts as well as precision farming. Various optical sensors can be attached to UAVs, each offering unique capabilities and sensitivities across different wavelengths of the electromagnetic spectrum. Recent development of UAV multispectral or hyperspectral imagery allows to simultaneously acquire both spectral and spatial information for a wide range of remote sensing studies (Aasen et al., 2015; Ballester et al., 2017; Caturegli et al., 2016; Potgieter et al., 2017; Wan et al., 2021; Xia et al., 2022; Zhou et al., 2021).

The outputs of spectral imagery, in the form of digital numbers (DNs), must undergo a series of radiometric correction processes to ensure accuracy and validity before being used to characterize plant physiological status and functional processes (Aasen et al., 2018). The initial steps of radiometric correction are to mitigate the dark signal noise, exposure discrepancies and vignetting effects, transforming the sensor outputs to normalized DN values which are uniform over the entire image throughout its operation (Beisl et al., 2010). On the other hand, spectral correction characterizes sensor's smile and keystone effects and ensures that the data accurately reflects the true spectral characteristics of the scene being observed (Jablonski et al., 2016). In some cases, absolute radiometric calibration is then performed to convert normalized DN values into radiance (Suomalainen et al., 2021). However, this conversion is not necessary if reflectance, rather than radiance, is desired in subsequent process. Spectral reflectance refers to the proportion of incident radiation reflected by a surface. As measurement of reflected radiation is key in the estimation of reflectance, accurate quantification of radiation received by the surface is of equal importance, especially in open-field environment where lighting condition is constantly changing.

In low-altitude UAV flights, solar irradiance can be strongly affected by dynamic environmental changes such as cloud cover, solar position and activity, and atmospheric absorption. Neglect to calibrate for the variability in incident solar irradiance would introduce substantial errors in the prediction of spectral reflectance (Aasen et al., 2018; Arroyo-Mora et al., 2021; Deng et al., 2018; Valencia-Ortiz et al., 2021). In spite of this, proper correction processes are often missing in real applications (Bergmüller and Vanderwel, 2022; K. C. Swain et al., 2010; Vega et al., 2015; Zaman-Allah et al., 2015; Zhang et al., 2018). Existing approaches to address this issue can be roughly grouped into three categories: 1) placing reference panels with known reflectance within the imaging region and performing the empirical line method (ELM) for correction, 2) taking advantage of an independent device collecting incident solar irradiance as reference, and 3) adopting pre-established models to normalize the influence from environmental condition changes.

The ELM is a common approach for reflectance calibration under stable illumination conditions for multispectral or hyperspectral imagery. A set of calibration reference panels, of which the reflectance were accurately measured in advance, were imaged before and/or after UAV flight missions, and a line is fitted with the least-squares method between each panel's DN values captured by the sensor and the ground truth reflectance (Smith and Milton, 1999). The derived relationship was then used to transform all the DN values in the imagery to the reflectance. Using multiple reference panels in line fitting reduces uncertainties (Lucieer et al., 2014; Wan et al., 2020), but simplified process can also be used where only one reference panel is imaged (Aasen et al., 2015; Cui et al., 2022). However, regardless of the number of reference panels, it requires them to be placed on the flat ground close to the imaging targets and away from shadows. While it might be managed with an effort in open areas with good accessibility, it imposes large restrictions of reflectance calibration in more complex terrains such as in a forest in hills with crowded vegetation and tall trees (Nevalainen et al., 2017). Furthermore, in order to achieve a reasonable reflectance calibration, the assumption of constant solar irradiance during the flight has to be met. It is usually recommended to schedule UAV spectral imagery at around noon on a clear sky day when the variation in light intensity is at its minimum and the UAV flight campaign is usually restricted within a short time window (Miura and Huete, 2009). The limitations significantly restrict its application scenarios and prevent its wide applications in agriculture, where a large proportion of agronomic traits are associated with specific growth stages and have to be evaluated in a high temporal resolution (Lyu et al., 2023).

Using an independent downwelling optical sensor to capture solar intensity and correct for light illumination variations during flight is another way for reflectance correction. The irradiance sensor can either be a broad-band downwelling light sensor (DLS) or a spectrometer, which is mounted on the top of the UAV looking upwards. Incident solar irradiance was captured in synchrony with spectral imaging sensors during the flight, allowing real-time reflectance generation without reflectance reference panels. So far, this method has been adopted in both academic communities and commercial products (MicaSense RedEdge; Waczak et al., 2024). However, experiments performed in a range of application scenarios and environmental conditions imply that the effectiveness of this method is less reliable, or at least questionable, compared against ground truth values or ELM methods (Barker et al., 2020; Daniels et al., 2023; DeCoffe et al., 2022; Deng et al., 2018; Franzini et al., 2019; Mamaghani and Salvaggio, 2019; Taddia et al., 2020; Wang, 2021). One possible cause can be attributed to the lack of a cosine corrector to measure the hemispherical irradiance (Olsson et al., 2021). Another possible cause may be due to the fact that the aircraft attitude is constantly changing during the flight, and the variation in aircraft orientations, especially pitch angle and roll angle, could introduce significant errors in sensor readings. To address this issue, fine adjustments in the design of cosine corrector and establishment of a tilt correction curve were carried out, and the error of irradiance measurements was managed to be controlled below 2% at solar zenith angles up to 70°, suggesting the necessity of careful thoughts in onboard sensor installation and data normalizations (Suomalainen et al., 2021). An alternative approach to capture dynamic solar irradiance is to have the second light sensor fixed on the ground facing downwards and constantly imaging a white panel placed beneath the spectrometer (Burkart et al., 2014; Xue et al., 2023). The solar spectrum measured on the reference panel was then used as reference for reflectance correction. Although the accuracy was proven in cases of small experimental sites, it remains to be further verified if this technique works on larger covering area as the distance between the ground-fixed sensor and imaging sensor gets substantially greater.

Additional efforts were put into the modelling of the relationship between reflectance estimates and environmental variations. For instance, by measuring the irradiance spectra of a grey panel at a continuous range of flight altitudes and utilizing the onboard spectrometer as reference, a calibration model was developed to eliminate the errors in reflectance estimates associated with atmospheric absorption (Suomalainen et al., 2021). A relative radiometric correction model was constructed based on concurrent satellite imagery, and demonstrated its capability in improving the radiometric consistency in multiflight UAV images (Jiang et al., 2022). Furthermore, diurnal visible and near-infrared hyperspectral change patterns for field-grown corn were explored based on high-frequency time-series monitoring, and a band-specific diurnal calibration model was proposed to reduce the variance in canopy reflectance caused by diurnal environmental fluctuations (Zhang et al., 2023). Although the modelling method required the least investment in hardware modifications, their effectiveness is only demonstrated in limited homogeneous environmental scenarios or datasets and might be prone to overfitting problems and accumulative error as more environmental factors came into play. Robustness of these models will need to be evaluated in further tests.

Although it is widely acknowledged that the heterogeneous environmental conditions introduce significant noise into multispectral imagery, a meta-analysis in UAV imagery studies showed only 38% of the published papers mentioned radiometric correction as part of the pre-processing stages in their research (Eskandari et al., 2020). It demonstrates the importance of raising awareness about correction methods in reflectance generation processing, and that methods in development need to take multidimensional factors into consideration, such as cost, adaptability and accuracy (Lu et al., 2020).

In this study, a new method utilizing an onboard, gimbal mounted downwelling spectrometer (DS) to correct the environmental impacts on the reflectance generation of a multispectral imager (MSI) was proposed. Firstly, we present the setup of the DS and the correction process associated, to ensure high-quality reference solar spectrum collection. Secondly, continuous synchronized spectrum collection between the DS and MSI was performed at different times and in various weather conditions to capture the heterogeneity in illuminations. Next, new models were constructed using features embedded in solar spectrum variations, and careful evaluations were conducted to compare them with conventional models across different environmental conditions. Finally, the performance of the proposed model was extensively validated on the UAV platform under various times and weather conditions to ensure prediction accuracy. This study aims to provide a reliable, cost-efficient and easy-to-operate method for reflectance generation, providing standardized data across varied illumination conditions and significantly expanding the time window for UAV multispectral remote sensing missions.

## 2. Materials and Methods

### 2.1 UAV imaging platform setup

An MSI and a DS were installed on a DJI M300 RTK UAV platform (Da-Jiang Innovations, Shenzhen, Guangdong, China; Fig. 1a). The UAV's upper surface, oriented skyward, was equipped with the DS, which was a miniature spectrometer (STS-NIR Spectrometer, Ocean Optics, Dunedin, USA) for measuring the downwelling solar spectrum. The DS recorded the solar spectrum within the range of 632–1,123 nm, with a spectral resolution of 6 nm. But, due to a low signal-to-noise ratio, data for wavelengths longer than 900 nm were excluded from the dataset. The DS was mounted on a two-axis upward gimbal, and was equipped with a cosine corrector to capture light across a 180° field of view. The upward gimbal was powered by a DC-DC voltage stabilizer module connected to the UAV. On the underside of the UAV, oriented towards the ground, multispectral imagery was acquired using a snapshot multispectral camera (MQ022HG-IM-SM5X5-NIR2, XIMEA GmbH, Muenster, Germany). The MSI was mounted on a three-axis downward gimbal, which enabled the capture of 25 spectral bands ranging from 600 to 875 nm in a single exposure with the image size of 409 × 216 pixels (Supplementary Table 1)(Geelen et al., 2015). The full width at half maximum of the spectral peaks varied between 10 nm to 15 nm. A band-pass filter with the range of 600 - 875 nm was installed on the lens. A spectral correction matrix provided by the sensor manufacturer was applied after reflectance generation.

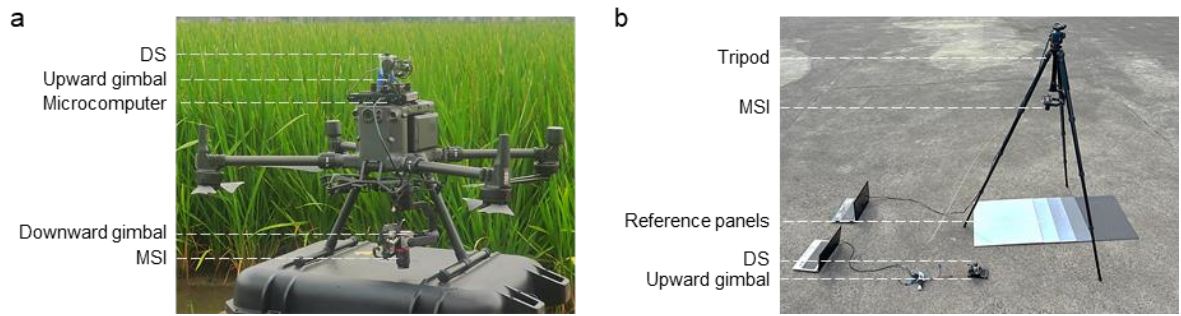

Fig. 1. Experimental setup for unmanned aerial vehicle (UAV) and ground imaging platforms with multispectral imager (MSI) and downwelling spectrometer (DS). (a) The customized UAV imaging platform enabled real-time acquisition of the solar spectrum via the onboard DS, and multispectral imaging of crop fields via the MSI. (b) The ground imaging platform also enabled real-time acquisition of solar spectrum with the DS, and multispectral imaging of five reference panels with reflectance of 7%, 15%, 30%, 50%, and 75%.

A microcomputer was mounted on top of the UAV to manage the triggering and data logging for both the DS and the MSI, as well as to synchronize timestamps between the two sensors. The total payload weight on the top of the UAV is approximately 610 grams, while the payload on the bottom is approximately 1020 grams. Sensor exposure time was adjusted based on sunlight intensity to prevent overexposure. Additionally, a customized wind shield was installed on the outer edge of the upward gimbal to minimize the impact of wind turbulence on the DS's measurement during flight (Supplementary Fig. S1). After installation, the gimbal required an average of approximately six seconds to re-stabilize (Supplementary Fig. S2). This was calculated by subtracting the visually detected time taken for the UAV to turn (~ 6 seconds) from the averaged time duration during which solar measurements were disturbed (~ 12 seconds).

## 2.2 Ground imaging platform setup

The same DS and MSI used in the airborne experiment were employed for the ground-based measurements (Fig. 1b). The MSI was securely mounted on a tripod, oriented vertically downward to capture multispectral images of five reference panels. The DS was positioned on a two-axis upward gimbal and placed adjacent to the reference panels. The exposure time for both devices was carefully monitored and adjusted as needed during the measurements to ensure that the DN values remained within 75% to 90% of the sensor's maximum range. Data acquisition was managed by two independent computers, which were positioned at a distance from the imaging area to prevent any potential shadows.

The five reference panels had nominal reflectance of 7%, 15%, 30%, 50%, and 75%, respectively (Changhui Electronic Technology Co., Guangzhou, Guangdong, China). Precise reflectance measurements of five reference panels were performed using an ASD FieldSpec 4 Spectroradiometer (Analytical Spectral Devices, Boulder, CO, USA), calibrated against a reference white panel of 99% reflectance made from Teflon (Jingyi Optoelectronics Technology Co., Ltd., Zhongshan, Guangdong, China). The manual measurements for these panels, in ascending order, averaged 8.4% (±0.21%), 17.0% (±0.27%), 31.6% (±0.47%), 51.3% (±0.24%), and 73.0% (±0.37%) within the wavelength range of 600-900 nm, indicating good uniformity (Supplementary Fig. S3). The same reference panels were used in UAV experiments.

## 2.3 Experimental design

### 2.3.1 Training data acquisition for reflectance prediction

To construct reflectance generation models, training data acquisition was conducted using the ground imaging system. Measurements of sunlight intensity and the reflectance of five reference panels were performed on three distinct days: a fluctuating cloudy day on February 27, 2024 (Modeling dataset 1), a sunny day on July 4, 2024 (Modeling dataset 2) and a partially cloudy day on June 24, 2024 (Modeling dataset 3). Multiple days of experiments were carried out to cover varying illumination conditions in the model. Data were continuously collected at 1-second intervals for the DS and at 30-second intervals for the MSI. The experiments were performed in an open area at Zhejiang University, Hangzhou, Zhejiang Province, China (30°17'51.00"N, 120°5'24.52"E), ensuring that measurements were free from shadows cast by surrounding buildings or vegetation.

### 2.3.2 UAV imaging campaigns across different time of the day

Twenty-six UAV flights were conducted throughout the day from sunrise to sunset with approximately 30-minute intervals on May 8, 2024. The weather was sunny with clear skies. The imaging campaign focused on breeding plots of Bok choy (*Brassica rapa* subsp. *chinensis*), with five reflectance panels placed in the center of the walkway. Due to a significant error introduced by the cosine corrector when the solar altitude was below 40°, only fifteen flights conducted between 08:30 and 15:30 were used for model evaluation. The UAV operated at an altitude of 15 m, with side and forward overlapping of 80%. The flight speed was 1.5 m/s, and each campaign covered an area of approximately 600 m$^2$, lasting around 10 min. Data from flights at 08:47 and 09:08 were excluded from the dataset due to overexposure in the multispectral imagery within the wavelength range of 649 nm to 679 nm, observed on the reference panel with reflectance of 75% (Supplementary Fig. S4). A breeding plot of Bok choy was manually labeled and extracted for the canopy reflectance.

### *2.3.3 UAV imaging campaigns under fluctuating cloudy weather*

A UAV imaging campaign was conducted from 12:25 to 13:08 on August 15, 2024. The imaging campaign focused on breeding plots of rice (*Oryza sativa*), with five reflectance panels placed on the edge of the field. The UAV operated at an altitude of 15 m, with flight speed of 1.5 m/s. Side and forward overlapping were set at 70%, which allowed for higher imaging efficiency and enabled coverage of a larger crop field area. The imaging area was approximately 12,700 m$^2$ and the total flight duration was 38 min. Reflectance and vegetation indices from a homogeneous crop field area were extracted for downstream data analysis.

### 2.4 Data preprocessing

### *2.4.1 Reference panel segmentation in multispectral imagery*

Due to the extensive size of the training data generated from the ground imaging system, an automated pipeline was developed to extract DN values of each reference panel from multispectral images. A significant challenge in this process was the shading caused by the tripod legs during continuous all-day measurements. Consequently, the area used for DN value extraction in the multispectral images needed to be dynamically adjusted to account for these shadings. To address this issue, the local mean and standard deviation were calculated within a sliding window, and a segmentation threshold was calculated as following:

$$T(x,y) = mean(x,y) - \rho * std(x,y) \tag{1}$$

where $mean(x,y)$ and $std(x,y)$ represent the local mean and local standard deviation calculated within a sliding window centered at pixel $(x,y)$. $\rho$ is a hyperparameter that needs to be tuned for optimal performance. Based on the preliminary test, a 15 × 15 sliding window and a value of 0.1 for hyperparameter $\rho$ was chosen based on the segmentation effectiveness. A shadow mask was then generated:

$$mask(x,y) = \begin{cases} 1 & if\ img(x,y) < T(x,y) \\ 0 & otherwise \end{cases} \tag{2}$$

where $mask(x, y)$ indicates whether a tripod shadow should be applied to the pixel $(x, y)$, while $img(x, y)$ represents the DN value of the multispectral images at that pixel. $T(x, y)$ denotes the segmentation threshold. Morphological operations were subsequently applied to refine the shadow mask and enhance the removal process. Subsequently, the average value of the unmasked area was calculated as the DN value of the reflectance panels. It needs to be mentioned that reference panel segmentations in UAV imagery were manually performed.

*2.4.2 Multispectral imagery preprocessing*

Following image acquisition, dark current correction and exposure normalization were performed. All images were normalized to a standard exposure time of 1 ms to ensure consistency across the dataset. The corrected DN values for the multispectral images were calculated as follows:

$$DN_{MSI,dark-corrected}(\lambda) = \frac{DN_{MSI,raw}(\lambda) - DN_{MSI,dark}(\lambda)}{t_{exp}} \quad (3)$$

where $DN_{MSI,raw}(\lambda)$ and $DN_{MSI,dark-corrected}(\lambda)$ represent the DN values for each band $\lambda$ in the multispectral imagery before and after corrections, respectively. $DN_{MSI,dark}(\lambda)$ denotes the dark current values for each band in the MSI, obtained by measuring the sensor readings with light completely blocked. $t_{exp}$ refers to the duration for which the multispectral images were exposed during capture.

Next, vignetting correction was performed to mitigate the darkening at the edges of images (Fig. 2). Under a stable lighting condition on a cloudy day, the MSI was fixed vertically to capture images of a 50% reflectance panel, which was rotated four times around its center axis, with three images taken in each of the four directions. The resulting 48 images were averaged to mitigate errors from uneven panel surfaces. A reference DN value for correction was obtained by averaging nine pixels at the center of the image. The vignetting correction coefficient for each pixel was then calculated using the following formula for each band:

$$DN_{MSI,normalized}(i, j, \lambda) = DN_{MSI,dark-corrected}(i, j, \lambda) * \frac{DN_{MSI,standard}(\lambda)}{DN_{MSI,panel}(i,j,\lambda)} \quad (4)$$

where $DN_{MSI,dark-corrected}(i, j, \lambda)$ and $DN_{MSI,normalized}(i, j, \lambda)$ represent the DN values for pixel $(i, j)$ of band $\lambda$ before and after vignetting correction, respectively. $DN_{MSI,standard}(\lambda)$ denotes the reference DN value derived from reflectance panel for band $\lambda$. $DN_{MSI,panel}(i, j, \lambda)$ denotes the measured DN value for pixel $(i, j)$ on reflectance panel for band $\lambda$.

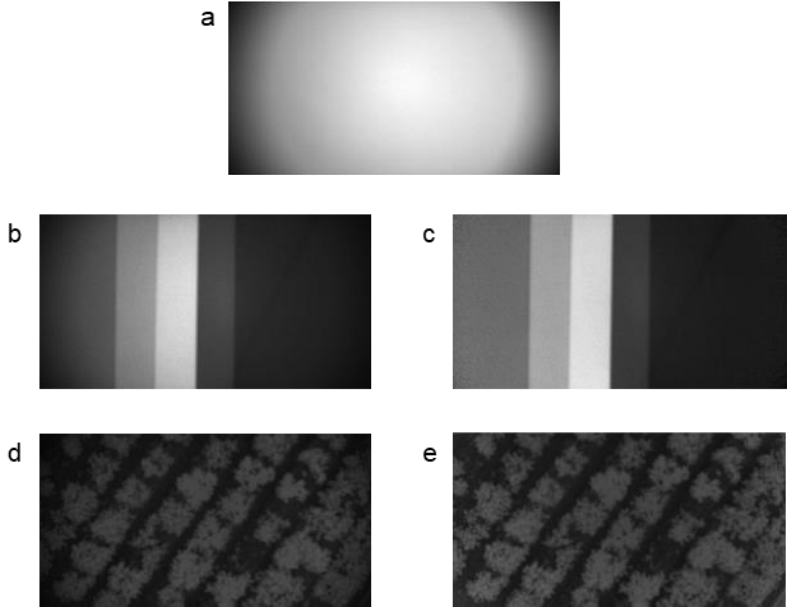

Fig. 2. Vignetting correction. (a) Reference panel with 50% reflectance was used to generate the vignetting correction file. Multispectral imagery of reference panels acquired in ground imaging platform (b) before and (c) after vignetting correction are shown. Multispectral imagery of crop field acquired in UAV imaging platform (d) before and (e) after vignetting correction are shown. The first band of the multispectral imagery at 603 nm is used for illustration.

*2.4.3 Downwelling spectrum preprocessing*

Similar to the multispectral imagery pre-processing, dark current correction and exposure normalization were performed. All spectra were normalized to a standard exposure time of 100 ms using the following equation:

$$DN_{DS,dark-corrected}(\lambda) = \frac{DN_{DS,raw}(\lambda) - DN_{DS,dark}(\lambda)}{t_{exp} / 100} \qquad (5)$$

where $DN_{DS,raw}(\lambda)$ and $DN_{DS,dark-corrected}(\lambda)$ represent the DN values of each band $\lambda$ produced by DS before and after corrections, respectively. $DN_{DS,dark}(\lambda)$ denotes the dark current readings for each band in the spectrum, obtained by completely covering the sensor head with a black, lightproof cloth. $t_{exp}$ refers to the duration for which the spectrum was exposed during measurements.

*2.4.4 Correction of downwelling spectrum for solar position*

On the same day and at the same location as described in section 2.3.2, UAV flights were conducted between the intervals of the imaging campaigns at each time points throughout the day. During each flight, the aircraft pilot manually performed two cycles of UAV rotations, with the DS continuously recorded the solar spectrum at 1-second intervals. Each flight was divided into multiple operations, with the UAV's heading angle rotated approximately 45° each time. Then, the aircraft was stabilized for 5 seconds, and the precise UAV heading angle and timing were recorded. This procedure was repeated until the UAV had completed two full rotations (720°). To mitigate background noise from solar intensity variations, the time series of DN values were smoothed using locally estimated scatterplot smoothing (LOESS). Time stamps were used to extract the solar altitude and azimuth angles corresponding to the solar spectrum at the points when the UAV headings were stabilized (Supplementary Fig. S5). Data from three flights were excluded from the analysis due to either low signal-to-noise ratios measured during the early morning or missing data resulting from manual errors.

The UAV-Sun angle, defined as the angle between the UAV heading angle and the solar azimuth angle, was calculated using the following equation:

$$\gamma_{UAV-Sun} = \gamma_{UAV} - \gamma_{Sun} \tag{6}$$

where $\gamma_{UAV}$ represents the UAV heading angle and $\gamma_{Sun}$ represents the solar azimuth angle. LOESS was used to remove the background noise in DN values caused by variations in solar intensity. The standardized deviation of DN value was calculated as:

$$\delta DN_{DS,dark-corrected}(\lambda) = \frac{DN_{DS,dark-corrected}(\lambda) - \overline{DN}_{DS,dark-corrected}(\lambda)}{\overline{DN}_{DS,dark-corrected}(\lambda)} \tag{7}$$

where $DN_{DS,dark-corrected}(\lambda)$ is the measured DN value and $\overline{DN}_{DS,dark-corrected}(\lambda)$ is the locally smoothed DN value of the DS for each band $\lambda$. A sinusoidal function was then fitted to the data as follows:

$$\delta DN_{DS,dark-corrected}(\lambda) \sim A * \cos(\alpha_{sun}) * \sin(\gamma_{UAV-Sun} + \varphi_0) \tag{8}$$

where $\delta DN_{DS,dark-corrected}(\lambda)$ represents the standardized deviation of DN value for each band $\lambda$. $\gamma_{UAV-Sun}$ is the UAV-Sun angle. $A$ and $\varphi_0$ were coefficients for amplitude and initial phase, respectively. $\alpha_{sun}$ denotes the solar altitude angle at the time of data measurement.

After acquiring the coefficients $A$ and $\varphi_0$, the predicted standardized DN value deviations derived from the sinusoidal model are denoted as $\delta \widehat{DN}_{DS,dark-corrected}(\lambda)$. Finally, the DN value of the DS, after normalization of each band $\lambda$, was calculated as follows:

$$DN_{DS,normalized}(\lambda) = DN_{DS,dark-corrected}(\lambda) - \delta \widehat{DN}_{DS,dark-corrected}(\lambda) * DN_{DS,dark-corrected}(\lambda) \tag{9}$$

The percentage of maximal DN value deviation, $P_{error}$, was calculated as follows:

$$P_{error} = A * \cos(\alpha_{sun}) * 2 \tag{10}$$

where $A$ represents the amplitude coefficient, and $\alpha_{sun}$ represents the solar altitude angle.

## 2.5 Reflectance generation models

Three classic methods for reflectance generation were constructed to draw comparisons with our method: ELM, direct correction method, and DLS method. In this study, we proposed principal component regression (PCR) models for reflectance generation, followed by a band selection process to reduce redundancy in the reference solar spectrum. Furthermore, multiple linear regression (MLR) was adopted to construct a lightweight model using selected sensitive bands in solar spectrum.

For model construction, Modeling dataset 1 (fluctuating cloudy weather) and Modeling dataset 2 (sunny weather) were combined and randomly split into training and testing sets with an 80% to 20% ratio, resulting in 5,208 and 1,302 samples per spectral band, respectively. Modeling dataset 3 (partial cloudy weather) and two UAV validation experiments on the Bok choy field and rice crop field were used to further evaluate the performance of the proposed model. Data for analysis were focused on solar altitude above 40°.

### 2.5.1 ELM

The following equation was used to convert DN values from multispectral imagery to reflectance (Baugh and Groeneveld, 2008):

$$R(i, \lambda) = DN(i, \lambda) * a_\lambda + b_\lambda \ (i = 1,2,3,4,5) \tag{11}$$

where $R(i, k)$ and $DN(i, k)$ denote the reflectance and DN values of the $i^{th}$ reference panel at band $\lambda$, respectively. $a_\lambda$ is the slope and $b_\lambda$ is the intercept.

For each multispectral image to be calibrated, two sets of reference data, each containing five reference panels, were used to determine $a_k$ and $b_k$. In practical applications of ELM, it is common to image reference panels before and after UAV imaging campaigns, with each imaging typically occurring at intervals of around 30 min. To replicate this setup in our experiment, one set of reference data was randomly sampled within 15 min prior to the multispectral image acquisition, while the other set was randomly sampled within 15 min following the multispectral image acquisition.

### 2.5.2 Direct correction method

Direct correction was applied to each synchronized multispectral image and solar spectrum data pair at specific bands. The ratio of DN values from the MSI and the DS was calculated using the following equation:

$$Ratio(\lambda) = DN_{MSI,normalized}(\lambda) / DN_{DS,normalized}(\lambda) \tag{12}$$

where $DN_{MSI,normalized}(\lambda)$ and $DN_{DS,normalized}(\lambda)$ denote the normalized DN values captured by the MSI and DS at band $\lambda$, respectively. A linear regression model was then fitted using following equation:

$$R(\lambda) = Ratio(\lambda) * a_\lambda + b_\lambda \tag{13}$$

where $R(\lambda)$ represents the reflectance at band $\lambda$. $a_\lambda$ is the slope and $b_\lambda$ is the intercept.

Since the three MSI bands in the range of 600 nm to 632 nm were not covered by the DS, these bands were excluded from the evaluation, resulting in twenty-two spectral bands for further analysis for direction correction method.

*2.5.3 DLS method*

Integration was performed on the DS solar spectrum to simulate a broadband light sensor. The ratio of DN values from the MSI and DS was calculated using the following equation:

$$Ratio(\lambda) = DN_{MSI,normalized}(\lambda) / \sum DN_{DS,normalized}(\lambda') \qquad (14)$$

where $DN_{MSI,normalized}(\lambda)$ denotes the normalized DN values captured by the MSI at band $\lambda$, and $DN_{DS,normalized}(\lambda')$ denotes the normalized DN values captured by the DS at band $\lambda'$. A linear regression model was then fitted using the following equation:

$$R(\lambda) = Ratio(\lambda) * a_\lambda + b_\lambda \qquad (15)$$

where $R(\lambda)$ represents the reflectance at band $\lambda$. $a_\lambda$ is the slope and $b_\lambda$ is the intercept.

*2.5.4 PCR method*

Ratio of DN values from the MSI and DS was calculated using the following equation:

$$Ratio(\lambda, \lambda') = DN_{MSI,normalized}(\lambda) / DN_{DS,normalized}(\lambda') \qquad (16)$$

where $DN_{MSI,normalized}(\lambda)$ denotes the normalized DN values captured by the MSI at band $\lambda$ and $DN_{DS,normalized}(\lambda')$ denote the normalized DN values captured by the DS at band $\lambda'$. Then, $Ratio(\lambda, \lambda')$ was transformed into a set of linearly uncorrelated PCs for each multispectral image at band $\lambda$. The PCs accounting for the maximum proportion of the total variance were utilized in the following PCR model to predict reflectance:

$$R(\lambda) = PC_1(\lambda) * a_\lambda^1 + PC_2(\lambda) * a_\lambda^2 + \cdots + PC_n(\lambda) * a_\lambda^n + b_\lambda \qquad (17)$$

where $R(\lambda)$ represents the reflectance at band $\lambda$. $PC_n(\lambda)$ denotes the $n^{th}$ PC (Principal Component) derived from PCA (Principal Component Analysis) for band $\lambda$. $a_\lambda^n$ is the coefficient associated with the $n^{th}$ PC for band $\lambda$, and $b_\lambda$ is the intercept for band $\lambda$.

*2.5.5 Band selection*

Following the calculation of $Ratio(\lambda, \lambda')$ as described in Section 2.5.4, integration over a 30 nm bandwidth was performed independently for each band $\lambda$, to simulate the response of a bandpass-filtered optical detector with a 30 nm bandwidth. The following equation was used in the calculation:

$$Ratio_{30nm}(\lambda, \lambda') = \int_{\lambda_1'}^{\lambda_2'} Ratio(\lambda, \lambda') \qquad (18)$$

where $Ratio(\lambda, \lambda')$ and $Ratio_{30nm}(\lambda, \lambda')$ represent the ratio values of the MSI band $\lambda$ and DS band $\lambda'$ before and after the integration, respectively. $\lambda_1'$ and $\lambda_2'$ define the limits of the 30 nm bandwidth, where $\lambda_2' - \lambda_1' = 30$ nm.

Subsequently, PCA was applied to $Ratio_{30nm}(\lambda, \lambda')$ for each band $\lambda$ to identify sensitive band candidates. To evaluate the significance of each band in model construction, MLR analyses were performed with various subsets of variables:

$$R(\lambda) = Ratio_{30nm}(\lambda, \lambda'_1) * a_\lambda^1 + Ratio_{30nm}(\lambda, \lambda'_2) * a_\lambda^2 + \cdots + Ratio_{30nm}(\lambda, \lambda'_n) * a_\lambda^n + b_\lambda \quad (19)$$

where $R(\lambda)$ represents the reflectance at band $\lambda$. $Ratio_{30nm}(\lambda, \lambda'_n)$ denotes the ratio value of the MSI and DS at $n^{th}$ candidate band after integration over a 30 nm bandwidth. $a_\lambda^n$ is the coefficient associated with the $n^{th}$ sensitive band candidate, and $b_\lambda$ is the intercept.

## 2.6 Analytical software and evaluation matrix

Multispectral imagery preprocessing was performed using MATLAB program (version 2021a, MathWorks Inc., America). For the smoothing and model fitting of DN values from DS during the correction for solar positions, the *loess* function and *nls* function in R (version 4.1.2) were employed. PCA and regression analysis were performed using *prcomp* and *lm* function in R. Scaling was enabled in PCA. Reflectance prediction and model evaluation used *terra* and *sf* package in R. The stitching of UAV multispectral images was executed using Agisoft Metashape (version 1.6.2, Agisoft LLC, St. Petersburg, Russia). The solar altitude angle and solar azimuth angle were calculated based on latitude, the Earth's declination angle, and the hour angle using MATLAB (Meeus, 1991).

The difference vegetation index (DVI) and normalized difference vegetation index (NDVI) were calculated as

$$DVI = NIR - R \quad (20)$$

$$NDVI = \frac{NIR - R}{NIR + R} \quad (21)$$

where R is the red band reflectance, and NIR is the near infrared band reflectance (Richardsons and Wiegand, 1977; Rouse et al., 1973). In this study, band at wavelength of 679 nm and 796 nm were used to represent red band and near infrared band, respectively.

RMSE and rRMSE were used to assess the accuracy of reflectance generation models. Calculations were performed for each MSI band $\lambda$ following these equations:

$$RMSE_\lambda = \sqrt{\frac{1}{N}\sum_{i=1}^{N}(R(\lambda) - \theta_\lambda)^2} \quad (22)$$

$$rRMSE_\lambda = \frac{RMSE_\lambda}{\overline{\theta_\lambda}} \quad (23)$$

where N is the sample number, $R(\lambda)$ and $\theta_\lambda$ are the predicted reflectance and ground truth reflectance for band $\lambda$. $\overline{\theta_\lambda}$ is the mean of the ground truth reflectance for band $\lambda$.

Coefficient of Variation (CV) was used to assess the relative variability of a dataset and it was calculated as following:

$$CV = \left(\frac{Standard\ Deviation}{Mean}\right) * 100\% \quad (24)$$

## 3. Results

### 3.1 Diurnal solar intensity variability and sensor correlations

Both the DN values recorded by the DS and measured on reference panels by the MSI exhibited significant variability across three distinct measurement days, each representing different weather conditions: a fluctuating cloudy day (Modeling dataset 1; Fig. 3a-b), a clear sunny day (Modeling dataset 2; Fig. 4a-b) and a partially cloudy day (Modeling dataset 3; Supplementary Fig. S6a-b). Modeling dataset 1 displayed pronounced fluctuations in cloudiness, leading to drastic changes in solar intensity across spectral bands. For instance, the average DN values measured by the DS increased from 2019.3 to 5154.2 between 11:46 and 11:50, making a 155.2% increase within just 4 min. When extending the time scale to 30 min, the maximum relative change in DN values reached 214.0% (e.g., from 12:49 to 13:09). In contrast, Modeling dataset 2 was characterized by predominantly stable sunny conditions, with some cloud cover occurring only around midday, and only a maximum change of 3.5% was observed over a 30-minute period (e.g., from 11:00 to 11:20). In Modeling dataset 3, intermittent thick clouds were observed around 10:00 and 14:00, with generally thinner clouds at other times, leading to minor yet continuous variations in solar intensity. During noon in Modeling dataset 2 and 3, despite the absence of visible cloud cover, DS measurements revealed solar intensity variations ranging from 8.9% to 15.4% within the same time periods. Solar intensity peaked around noon each day, with the daily maximum intensity approximately 33% higher in Modeling datasets 2 and 3 compared to Modeling dataset 1, which can attributable to seasonal variations.

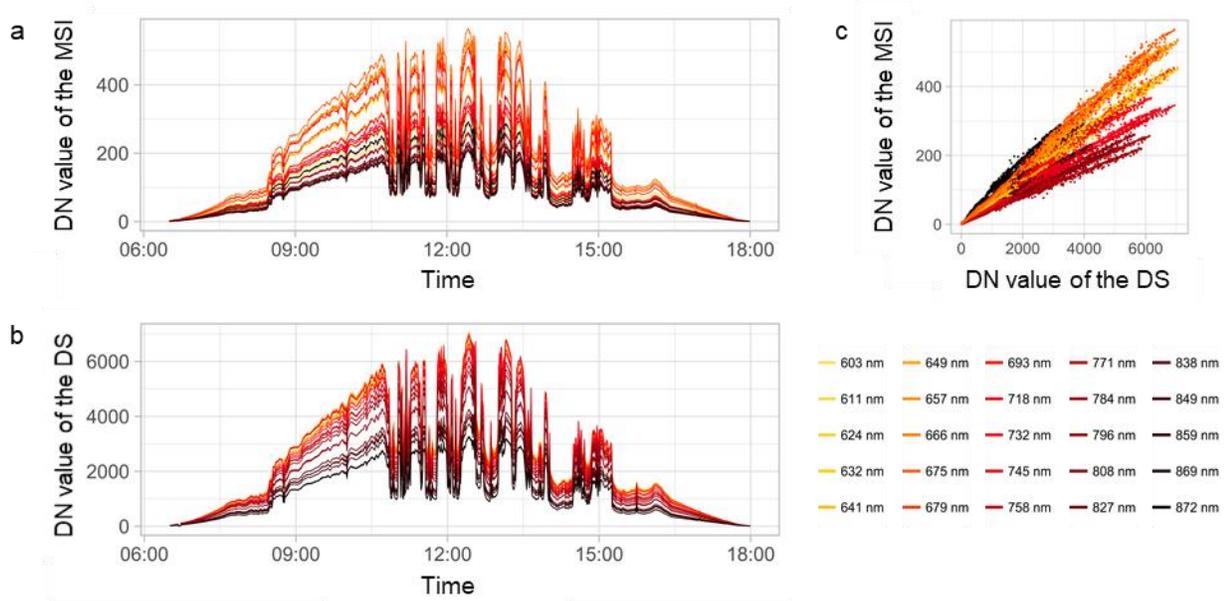

Fig. 3. Variation in solar intensity on February 27, 2024, exhibiting substantial fluctuations due to changing cloud cover. This variability is reflected in the DN (digital number) readings from both the (a) multispectral imager (MSI) and (b) downwelling spectrometer (DS). (c) The correlations between readings from the two sensors are shown for each band. Data acquired from the MSI on the 75% reflectance panel are used for illustration. For DS, the bands closest to each of the MSI bands are shown.

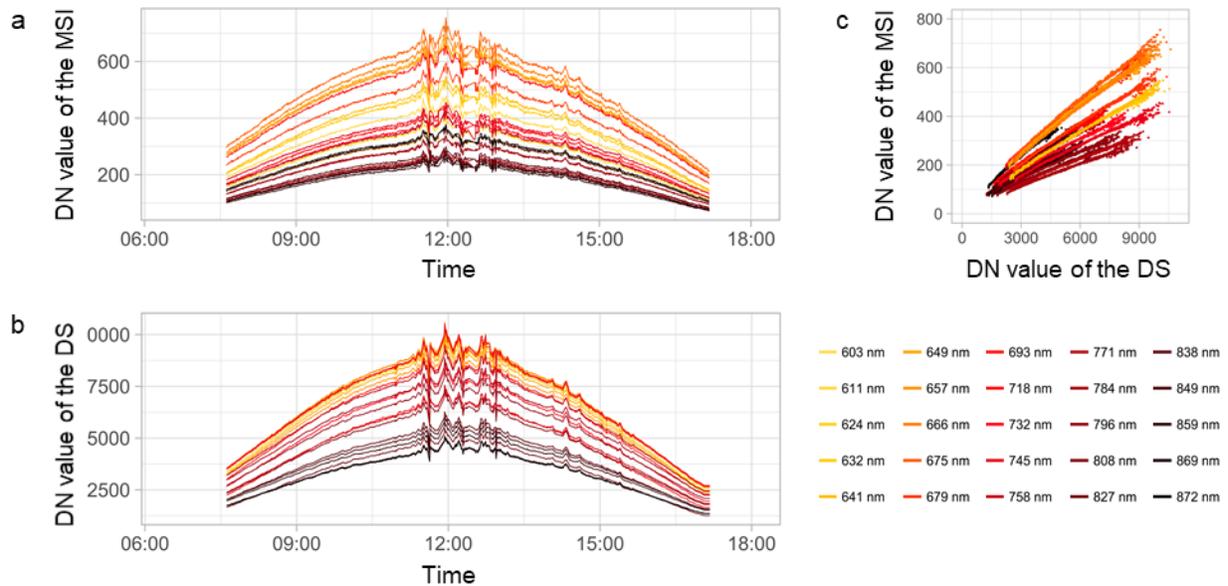

Fig. 4. Variation in solar intensity on July 4, 2024, exhibiting predominantly stable sunny weather. This variability is reflected in the DN (digital number) readings from both the (a) multispectral imager (MSI) and (b) downwelling spectrometer (DS). (c) The correlations between readings from the two sensors are shown for each band. Data acquired from the MSI on the 75% reflectance panel are used for illustration. For DS, the bands closest to each of the MSI bands are shown.

DN values measured on reference panels by the MSI exhibited strong correlations with the DN values output by the DS, regardless of illumination conditions, although the ratios of these two readings varied across bands (Fig. 3c; Fig. 4c; Supplementary Fig. S6c). For example, examining the 75% reflectance reference panel in Modeling dataset 1, ratios between DN values from the MSI and DS ranged from 0.037 to 0.085 for bands with central wavelengths of 771 nm and 859 nm, respectively. Overall, the correlations between corresponding bands of MSI and DS had an average $R^2$ of 0.89 (p<0.001), indicating synchronized measurements between the sensors. Ratios of DN values from the MSI and DS decreased on panels with lower reflectance, but the high correlation between the two sensors was constantly remained (Supplementary Fig. S7).

## 3.2 Correction of DS measurements for solar position

The DS employed in this study was mounted on a gimbal and shielded from the wind to ensure stability, but errors were still introduced, probably due to the unevenness of the diffusing plate. This is evident by plotting the deviation in DN values of the DS as a function of UAV-Sun angles—defined as the angles between the UAV heading and the solar azimuth angle (Fig. 5a). It was observed that the error in DN values varied with changes in the UAV-Sun angle and solar altitude angle (Fig. 5a). Calculated from the amplitude of each DN value deviation curve for flights conducted at different times of the day with varying solar altitude angles, the percentage variation in maximum DN values displayed a symmetrical pattern relative to the time of day, with errors peaking in the early morning and late afternoon when the solar altitude angle was smallest (Fig. 5c). Although the DN values varied by only approximately 2.5% when UAV flights were conducted at different heading angles at noon, this error increased to 10% at 09:00 in the morning and 15:00 in the afternoon, when the solar altitude angle was around 50°. The error further escalated to over 40% in the late afternoon when the solar altitude angle dropped to 10°. A linear relationship between the error and the cosine of solar altitude angle was observed for cosine values no exceeding 0.75 (Fig. 5d). On the day of experiment, the solar altitude angle was 27.0° at the start of the experiment in the morning, peaked at 81.3° around noon, and decreased to 9.2° in the late afternoon (Fig. 5e). The threshold of cosine of solar altitude angle at 0.75 corresponded to a solar altitude angle of approximately 40°. Additionally, the observed DN value curves exhibited characteristics consistent with a sinusoidal function, specifically showing periodic oscillations with a regular, smooth waveform. This sinusoidal pattern became more apparent after standardizing the DN deviations to a common scale (Fig. 3f). The curves had a frequency of 1 and a fixed initial phase, indicating that the function completed one full cycle over an interval of 360° and oscillated in synchrony with the UAV-Sun angle. The amplitude of the sinusoidal function was modeled as the product of a fitting coefficient, termed amplitude factor, and the cosine of the solar altitude angle. After fitting the DN values across flights to a sinusoidal function based on the UAV-Sun angle, and averaging estimates across bands, the averaged amplitude factor was -5.7, while the averaged initial phase was -15.0° (Fig. 5g-h). Finally, the DN values of the DS were adjusted by subtracting the predicted deviations from the original measurements for all bands within the measurement range (Fig. 5i).

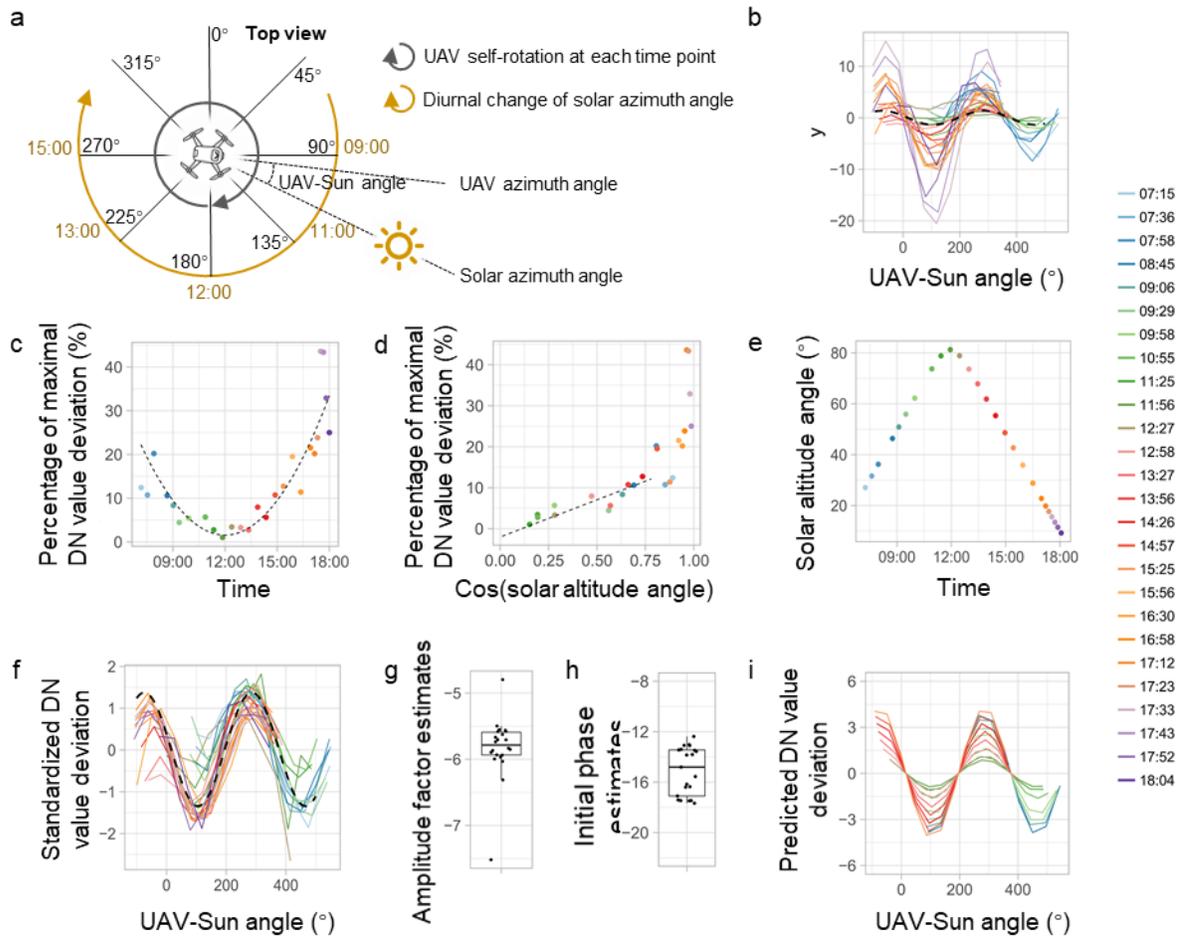

Fig. 5. Correction for the downwelling solar intensity measurements. (a) Two full cycles of UAV rotations were performed in different time points in the day with continuous downwelling spectrometer (DS) measurements. (b) Changes in DN (digital number) value deviations were observed in response to variations in the UAV-Sun angle across different time points of the day. The percentage of maximal DN value deviation is plotted against (c) time of day and (d) cosine of solar altitude angle. (e) The solar altitude angle throughout the day are shown. (f) After standardizing the DN value deviations, a sinusoidal trend was detected. The (g) amplitude factor and (h) initial phase coefficients for the sinusoidal function were estimated through model fitting. (i) Model predictions of DN value deviation are plotted. Dashed lines in (b) and (f) indicate reference sinusoidal functions. Dashed lines in (c) and (d) indicate a quadratic function and a linear function fitted to the data points, respectively. Only flights with a solar altitude angle above 40° are plotted in (i). For illustration, band at 632 nm from the DS is used in the plotting.

The performance of the correction model was subsequently assessed using DS measurements from the Modeling datasets. Changes in the solar azimuth angle relative to the stationary position of the DS led to noticeable discrepancies between morning and afternoon DN measurements when comparing the DN values on five reference panels from the MSI with those from the DS (Fig. 6a). The application of the correction model successfully corrected these time-related errors in the DN values (Fig. 6b). The rRMSE calculated between the scaled DN values of the MSI and the DS indicated a maximum error reduction of 41.0% following the implementation of the correction model, with this improvement observed on the 30% reflectance panel.

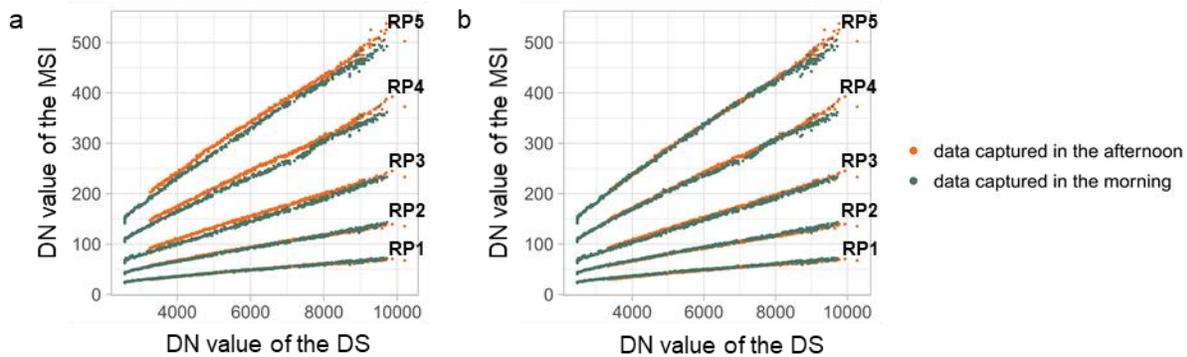

Fig. 6. Effectiveness of the solar intensity correction model. (a) Before correction, the correlations of DN (digital number) measurement from the downwelling spectrometer (DS) and the multispectral imager (MSI) were notably different for data taken in the morning and in the afternoon. (b) The differences were eliminated after the correction. For illustration, bands at 632 nm from the multispectral imager and the DS are used in the plotting. Reference panels (RPs), labeled RP1 through RP5, indicate panels with reflectance levels of 7%, 15%, 30%, 50%, and 75%, respectively.

## 3.3 PCA and effective band selection on downwelling solar spectrum

Modeling dataset 1 and Modeling dataset 2, representing significant solar spectral variance for cloudy and sunny weather conditions respectively, were combined for PCA and subsequent modeling. After standardizing the DN values from the DS in these datasets, substantial variance was observed in the spectral ranges of 715 nm to 730 nm, 758 nm to 762 nm, and 813 nm to 823 nm (Fig. 7a). PCA was performed on the ratio of DN values from the reflectance panels measured by the MSI to the DN values measured by the DS. The first four PCs accounted for 99.9% of the total variance in the dataset (Fig. 7b-e). Due to high intercorrelation among the solar spectrum bands, the loadings in PC1 were relatively uniform across the spectrum. In contrast, wavelengths at 718 nm, 727 nm, 762 nm, 814 nm, 898 nm, and 914 nm exhibited the most significant loadings in PCs 2 through 4. These wavelengths align with the atmospheric absorption features of water vapor and oxygen gas, highlighting their considerable influence on the solar spectrum in low-altitude remote sensing across different times and environmental conditions. The top PCs were then used to develop MLR models, with one to four PCs included in each model, and their performance was evaluated in the next section.

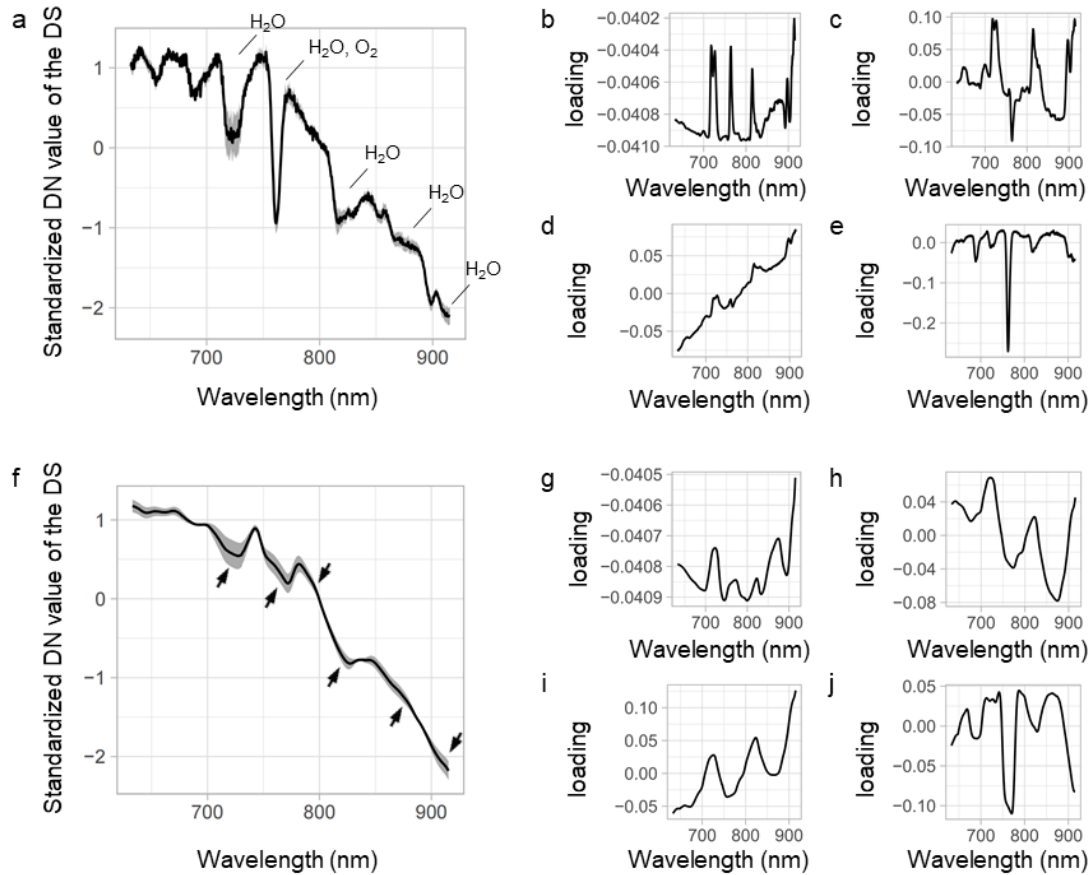

Fig. 7. Principal component analysis (PCA) results of the downwelling solar spectrum. (a) After standardization, the downwelling spectrometer DN (digital number) values exhibited the most variance within wavelength ranges sensitive to atmospheric absorption by water vapor and oxygen gas, as highlighted in the annotations. (b-e) Loadings for the top four principal components (PCs) derived from the PCA are shown. The spectra were then integrated over a 30 nm bandwidth to reduce the spectral resolution. (f) Following this transformation, the standardized DN values closely resembled the original spectra, and (g-j) loadings for the top four PCs are shown. The wavelength with significant loadings are indicated by arrows in (f).

To investigate the feasibility of using narrow-band detectors as substitutes for full-spectrum spectrometers to lower correction costs, we performed transformations by integrating each spectrum over a 30 nm bandwidth, assuming the use of an optical filter with a 30 nm bandwidth. The standardized DN values from the DS and the PCA results after this transformation closely resembled the original spectrum in terms of wavelength ranges with significant variance (Fig. 7f). However, the peaks and troughs in the spectrum became less pronounced. Wavelengths at 722 nm, 773 nm, 800 nm, 822 nm, 874 nm, and 915 nm exhibited the highest loadings in each of the principal components and were therefore selected as candidate wavelengths for optimal band combinations (Fig. 7g-j).

MLR analyses were conducted for each MSI band, utilizing all possible combinations of six candidate bands mentioned above, and the models were evaluated under two different weather conditions (Table 1). As expected, model accuracy was superior in sunny conditions compared to fluctuating cloudy conditions, with RMSE and rRMSE values being, on average, 35.6% and 19.1% lower in sunny weather, respectively. Model error generally decreased with the inclusion of more variables, although the rate of improvement diminished with additional variables. For the top-performing one-band models, RMSE and rRMSE averaged 2.60% and 8.2%, respectively. In contrast, the six-band model achieved RMSE and rRMSE values of 1.38% and 5.4%, reflecting improvements of 47.1% and 33.6%, respectively. Upon evaluating each category with different numbers of included bands, the best-performing sets were identified as follows: {722 nm}, {773 nm, 822 nm, 915 nm}, {722 nm, 773 nm, 800 nm, 915 nm}, {722 nm, 773 nm, 822 nm, 874 nm, 915 nm}, and {722 nm, 773 nm, 800 nm, 822 nm, 874 nm, 915 nm}. Averaged across both weather conditions, the corresponding RMSE values were 2.30%, 1.79%, 1.55%, 1.41%, 1.37%, and 1.38%, and rRMSE values were 7.4%, 6.4%, 5.7%, 5.5%, 5.4%, and 5.4%, respectively. Among these models, the four-band combination {722 nm, 773 nm, 800 nm, 915 nm} balanced model performance and the number of bands required, with RMSE and rRMSE values of 1.76% and 6.1% in fluctuating cloudy weather and 1.07% and 4.8% in sunny weather. Thus, this 4-Band MLR model was chosen for further analysis in the next section.

Table 1. Best performing multiple linear regression models using varying number of wavelength as predictors, chosen from six sensitive bands (722 nm, 773 nm, 800 nm, 822 nm, 874 nm and 915 nm). Each predictor was transformed to the integral of the central wavelength of λ and bandwidth = 30 nm.

| Band number | Fluctuating cloudy weather | | | Sunny weather | | |
|---|---|---|---|---|---|---|
| | Set of bands (λ) | RMSE (%) | rRMSE (%) | Set of bands (λ) | RMSE (%) | rRMSE (%) |
| 1 Band | {722 nm} | 2.65 | 7.6 | {722 nm} | 1.94 | 7.2 |
| | {822 nm} | 3.03 | 8.7 | {915 nm} | 1.99 | 6.7 |
| | {915 nm} | 3.48 | 10.3 | {822 nm} | 2.51 | 8.4 |
| 2 Bands | {722 nm nm, 773 nm} | 2.01 | 6.6 | {773 nm nm, 800 nm} | 1.48 | 5.7 |
| | {722 nm nm, 800 nm} | 2.14 | 6.7 | {722 nm nm, 773 nm} | 1.54 | 6.0 |
| | {722 nm nm, 874 nm} | 2.14 | 6.7 | {800 nm nm, 874 nm} | 1.58 | 6.2 |
| 3 Bands | {773 nm, 822 nm, 915 nm} | 1.84 | 6.3 | {773 nm, 800 nm, 915 nm} | 1.02 | 4.6 |
| | {722 nm, 773 nm, 874 nm} | 1.93 | 6.5 | {773 nm, 822 nm, 915 nm} | 1.20 | 5.1 |
| | {722 nm, 773 nm, 800 nm} | 1.95 | 6.5 | {773 nm, 800 nm, 822 nm} | 1.25 | 5.0 |
| 4 Bands | {722 nm, 773 nm, 800 nm, 915 nm} | 1.76 | 6.1 | {722 nm, 773 nm, 874 nm, 915 nm} | 0.99 | 4.7 |
| | {722 nm, 773 nm, 800 nm, 822 nm} | 1.76 | 6.1 | {722 nm, 773 nm, 822 nm, 874 nm} | 1.01 | 4.8 |
| | {722 nm, 773 nm, 822 nm, 874 nm} | 1.78 | 6.2 | {722 nm, 773 nm, 800 nm, 915 nm} | 1.07 | 4.8 |
| 5 Bands | {722 nm, 773 nm, 822 nm, 874 nm, 915 nm} | 1.75 | 6.1 | {722 nm, 773 nm, 800 nm, 874 nm, 915 nm} | 0.99 | 4.7 |
| | {722 nm, 773 nm, 800 nm, 874 nm, 915 nm} | 1.76 | 6.1 | {722 nm, 773 nm, 800 nm, 822 nm, 874 nm} | 0.99 | 4.7 |
| | {722 nm, 773 nm, 800 nm, 822 nm, 915 nm} | 1.77 | 6.1 | {722 nm, 773 nm, 822 nm, 874 nm, 915 nm} | 1.00 | 4.7 |
| 6 Bands | {722 nm, 773 nm, 800 nm, 822 nm, 874 nm, 915 nm} | 1.75 | 6.1 | {722 nm, 773 nm, 800 nm, 822 nm, 874 nm, 915 nm} | 1.00 | 4.7 |

### 3.4 Model evaluations and comparisons

Performance evaluations were conducted using RMSE and rRMSE, averaged across all MSI bands under varying cloudy and sunny conditions. This included the proposed PCR models, the 4-Band MLR model, and conventional methods of ELM, direct correction, and DLS (Fig. 8a-b). Evaluated across weather conditions, the direct correction model, Top-4PC PCR model, and 4-Band MLR model showed the best performance, with average RMSE values of 1.36%, 1.40%, and 1.37%, and average rRMSE values of 5.3%, 5.4%, and 5.4%, respectively. Tukey's test revealed no statistically significant differences among these three models (p = 0.001). The Top-3PC PCR, Top-2PC PCR, and Top-1PC PCR models followed, with average RMSE values of 1.85%, 1.86%, and 3.38%, and average rRMSE values of 6.2%, 6.3%, and 10.1%, respectively. Incorporating the second PC into the Top-1PC PCR model resulted in a notable 44.7% improvement in average RMSE. But, beyond this addition, the model exhibited varying sensitivity to illumination conditions. The DLS model yielded RMSE and rRMSE values of 3.37% and 10.1%, which were close to the Top-1PC PCR model's performance. In comparison with DLS model, the RMSE and rRMSE of 4-Band MLR model were reduced by 59.5% and 46.6%, respectively, averaged across weather conditions. The ELM model showed the highest overall error but exhibited distinct performance variations across different weather conditions. Under sunny conditions, the ELM model achieved unexpectedly low RMSE and rRMSE values of 1.21% and 4.6%, outperforming the DLS model and all PCR models except for Top-4PC. However, the ELM model's performance deteriorated significantly under fluctuating cloudy weather, with errors increasing by an order of magnitude. In comparison with ELM model, the RMSE and rRMSE of 4-Band MLR model were reduced by 17.9% and 0.2%, respectively under sunny weather, and by 86.1% and 83.6%, respectively under fluctuating cloudy weather. Notably, all correction methods performed equally well for MS bands both inside and outside the wavelength range of the DS spectra, except for the direct correction method, which was not feasible in principle (Table 2).

Given that the 4-Band MLR model performed best among all models and offers cost advantages by replacing a spectrometer with four narrow-band sensors, it was proposed as a superior reflectance generation method, and underwent further validation. The 4-Band MLR model was used to predict reflectance for five reference panels on the test set of Modeling dataset 1 (Fig. 8c) and Modeling dataset 2 (Fig. 8d), and an independent validation set of Modeling dataset 3 (Fig. 8e). Despite dynamic changes in solar position between 09:00 and 15:00 across different seasons and weather conditions, the stability of the reflectance in 25 bands of the MSI was largely maintained in all datasets. The average RMSE and rRMSE values for reflectance predictions in these datasets were 1.79% and 6.1%, 1.07% and 4.8%, and 2.24% and 7.2%, respectively.

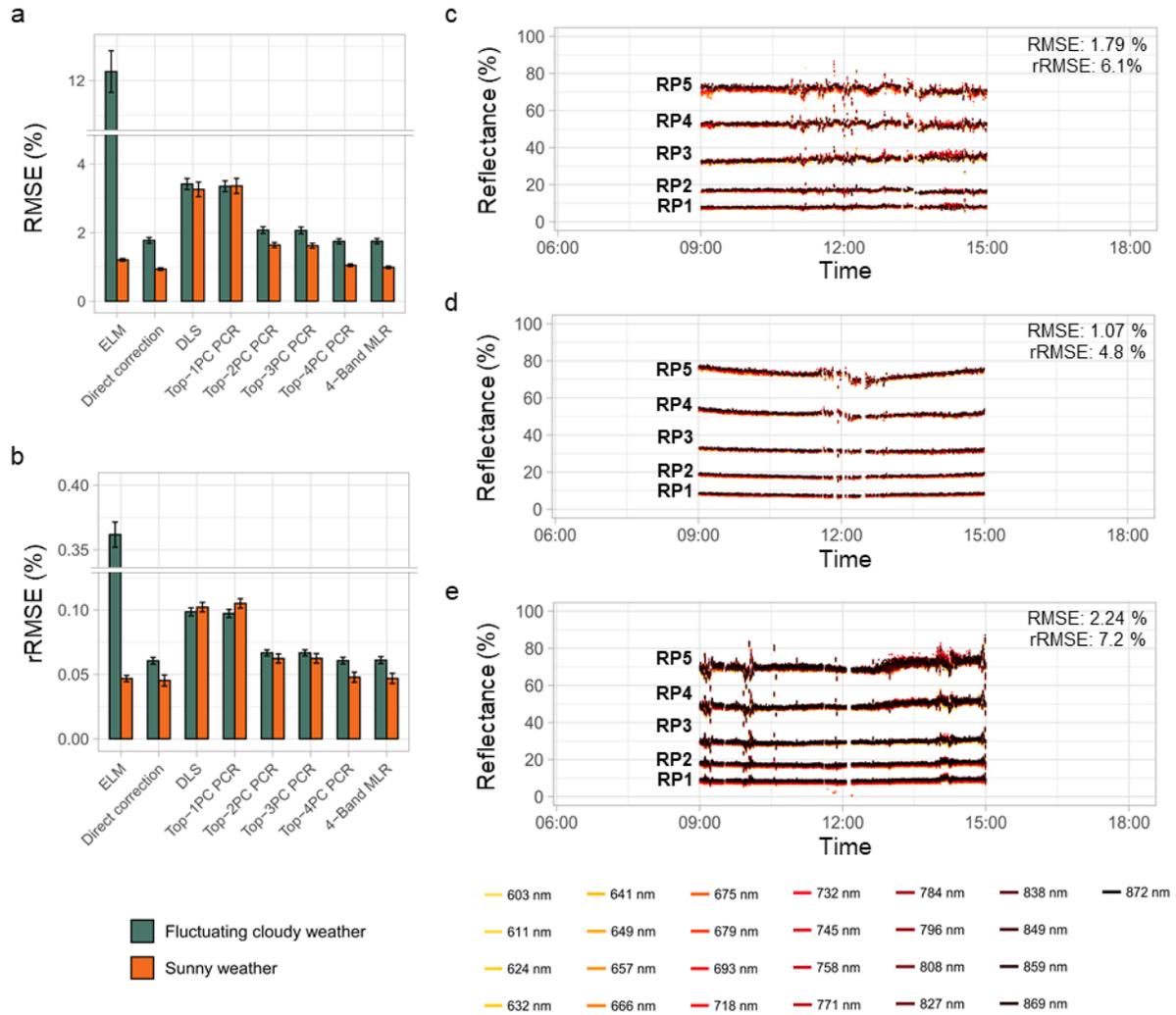

Fig. 8. Performance comparisons and evaluations for various reflectance generation models. Performance of the proposed Top-1PC, Top-2PC, Top-3PC, and Top-4PC principal component regression (PCR) models and the 4-Band multiple linear regression (MLR) model were evaluated and compared with the conventional methods of empirical line method (ELM), the direct correction model, and the downwelling light sensor (DLS) model, under different weather conditions. (a, b) Statistical metrics of RMSE and rRMSE were utilized for evaluation. Reflectance predictions were generated using the 4-Band MLR model on different datasets: (c) the test dataset of Modeling dataset 1, characterizing fluctuating cloudy weather, (d) the test dataset of Modeling dataset 2, characterizing sunny weather, and (e) an independent validation dataset from Modeling dataset 3, characterizing partial cloudy weather. Reference panels (RPs), labeled RP1 through RP5, indicate panels with reflectance levels of 7%, 15%, 30%, 50%, and 75%, respectively.

Table 2. Statistical summary of each model performance under fluctuating cloudy day and sunny day weather.

| | MSI band 22-25 (inside wavelength range of DS spectra) | | | | MSI band 1-3 (outside wavelength range of DS spectra) | | | |
|---|---|---|---|---|---|---|---|---|
| weather condition | method | RMSE (%) | rRMSE (%) | weather condition | method | RMSE (%) | rRMSE (%) |
| fluctuating cloudy day | ELM | 12.59 | 36.9 | fluctuating cloudy day | ELM | 9.87 | 31.1 |
| | Direct correction | 1.77 | 6.1 | | Direct correction | / | / |
| | DLS | 3.44 | 9.9 | | DLS | 3.25 | 9.4 |
| | Top-1PC PCR | 3.36 | 9.8 | | Top-1PC PCR | 3.27 | 9.5 |
| | Top-2PC PCR | 2.07 | 6.6 | | Top-2PC PCR | 2.08 | 6.9 |
| | Top-3PC PCR | 2.07 | 6.6 | | Top-3PC PCR | 2.03 | 6.9 |
| | Top-4PC PCR | 1.74 | 6.0 | | Top-4PC PCR | 1.76 | 6.4 |
| | 4-Band MLR | 1.75 | 6.1 | | 4-Band MLR | 1.73 | 6.5 |
| sunny day | ELM | 1.21 | 4.6 | sunny day | ELM | 1.19 | 4.9 |
| | Direct correction | 0.94 | 4.5 | | Direct correction | / | / |
| | DLS | 3.30 | 10.2 | | DLS | 3.02 | 10.1 |
| | Top-1PC PCR | 3.39 | 10.5 | | Top-1PC PCR | 3.17 | 10.6 |
| | Top-2PC PCR | 1.64 | 6.2 | | Top-2PC PCR | 1.62 | 6.5 |
| | Top-3PC PCR | 1.62 | 6.2 | | Top-3PC PCR | 1.60 | 6.5 |
| | Top-4PC PCR | 1.05 | 4.7 | | Top-4PC PCR | 1.03 | 5.1 |
| | 4-Band MLR | 0.99 | 4.6 | | 4-Band MLR | 0.97 | 5.0 |

### 3.5 Validation on UAV imaging campaign at different time throughout a sunny day

To validate the performance of the 4-Band MLR model in field experiments, 15 imaging campaigns were conducted to acquire crop field multispectral images at various times of the day with varying solar positions. The original DN values of the multispectral images measured on the reference panels exhibited significant variance associated with spectral bands and imaging time (Fig. 9a). For example, DN values for the wavelength of 675 nm ranged from 74.3 to 110.9, 121.4 to 219.2, 195.3 to 344.3, 304.7 to 497.8, and 424.4 to 686.9 for each reference panel, respectively, in ascending order of reflectance values. The average CVs across the bands were calculated to be 13.2%, 16.8%, 16.7%, 15.7%, and 12.6% for each reference panel, indicating significant variations in solar intensity at different times of the day.

Following the generation of reflectance, the influence of environmental factors on the data was notably reduced (Fig. 9b). For the wavelength of 675 nm, the reflectance ranged from 5.9% to 8.9%, 15.7% to 20.9%, 29.1% to 35.3%, 47.3% to 55.8%, and 66.0% to 78.8% for each reference panel, respectively. The average CVs across the bands were calculated as 9.6%, 7.3%, 6.2%, 5.0%, and 5.6% for each reference panel, resulting in a 55.1% reduction in variation. The averaged RMSE and rRMSE across bands were 0.82% and 9.0%, 1.21% and 7.1%, 2.00% and 6.3%, 2.46% and 4.8%, and 3.66% and 5.0% for each reference panel, respectively, again in ascending order of reflectance values. Overall, the average RMSE and rRMSE were 2.03% and 6.6%, demonstrating that the stability of the reflectance across all twenty-five multispectral bands was largely maintained, despite the images being captured at different times of the day.

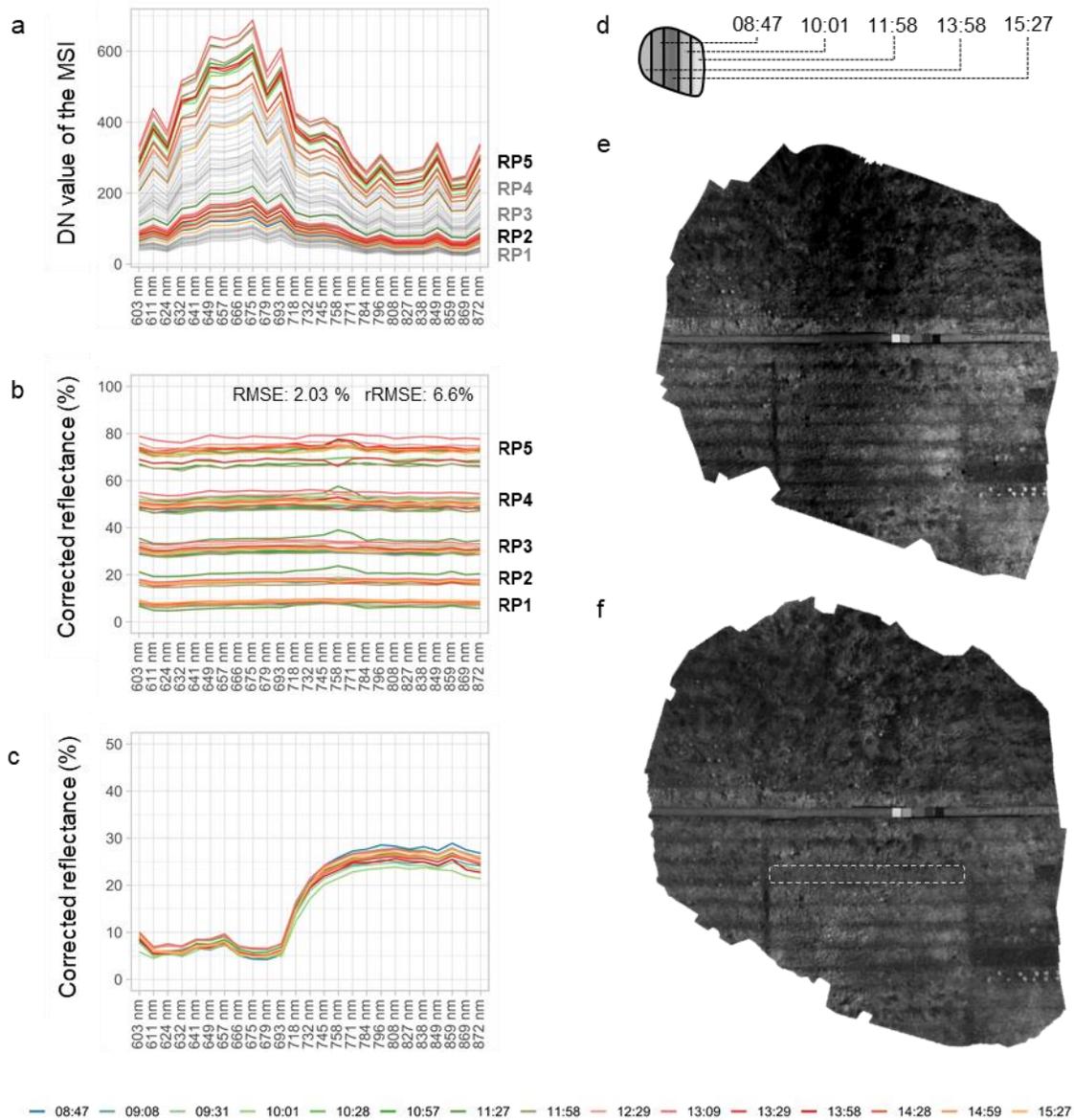

Fig. 9. Model validation in UAV imaging campaigns at different times throughout a sunny day. (a) Original DN (digital number) values are shown for each spectral band collected by the MSI on five reference panels at various times of day. Using the 4-Band MLR model, reflectance was predicted for the same image set. Reflectance for (b) the five reference panels and (c) a nearby vegetation area are displayed cross time points. (d) Additionally, images captured from five distinct time points were merged to create an orthomosaic image, with each time point represented as a vertical stripe within the composite area. Comparisons can be made between (e) the orthomosaic image created with raw DN values and (f) the one generated after converting the images to reflectance. The vegetation area is labeled with dashed rectangle in (f). For illustration, the band at wavelength of 827 nm of the multispectral imagery is displayed in greyscale. Reference panels (RPs), labeled RP1 through RP5, indicate panels with reflectance levels of 7%, 15%, 30%, 50%, and 75%, respectively.

In the evaluation of the spectral measurements obtained from vegetation, canopy reflectance was extracted from a relatively homogeneous Bok choy breeding plot, revealing consistent spectral measurements across different times of the day (Fig. 9c). For the bands adjacent to the red edge, the reflectance ranged from 4.22% to 6.63% at wavelength of 679 nm and from 21.46% to 25.88% at wavelength of 758 nm. The averaged CV across bands was calculated to be 8.2%. The observed errors in the measurements may arise from a combination of prediction errors, variability in canopy structure between imaging campaigns, and the plant's circadian responses to environmental conditions.

To visually demonstrate the drastic lighting differences in images captured at different times of the day, an orthomosaic image was created by merging images from five distinct time points. Each time point contributed a vertical stripe to the overall area (Fig. 9d). Without any corrections, the orthomosaic image exhibited a noticeable difference between brighter and darker areas. This discrepancy was primarily due to significant variations in DN values recorded by the MSI under different environmental conditions (Fig. 9e). In contrast, when all images were converted to reflectance using the 4-Band MLR model, and the same process was repeated for orthomosaic image creation, no visible differences in regional brightness were observed, resulting in a more uniform representation of the area across the different time points (Fig. 9f).

### 3.6 UAV imaging campaign validations under fluctuating cloudy weather conditions

To further validate the precision of the proposed reflectance generation method, an imaging campaign was conducted over a large crop field on a day characterized by significant fluctuations in cloud cover. An orthomosaic was constructed using multispectral images with reflectance correction based on reference panels via the ELM (Fig. 10a). This was compared to an orthomosaic constructed using images generated with the 4-Band MLR model (Fig. 10b). The ELM-corrected orthomosaic exhibited significant variation in reflectance along the UAV flight trajectory, attributed to continuous changes in environmental lighting conditions due to dynamic cloud movements. In contrast, this variation was largely mitigated when the 4-Band MLR model was employed. This demonstrates that the dynamic fluctuations in the solar spectrum were captured by the DS in real-time, providing a valuable reference for reflectance normalization.

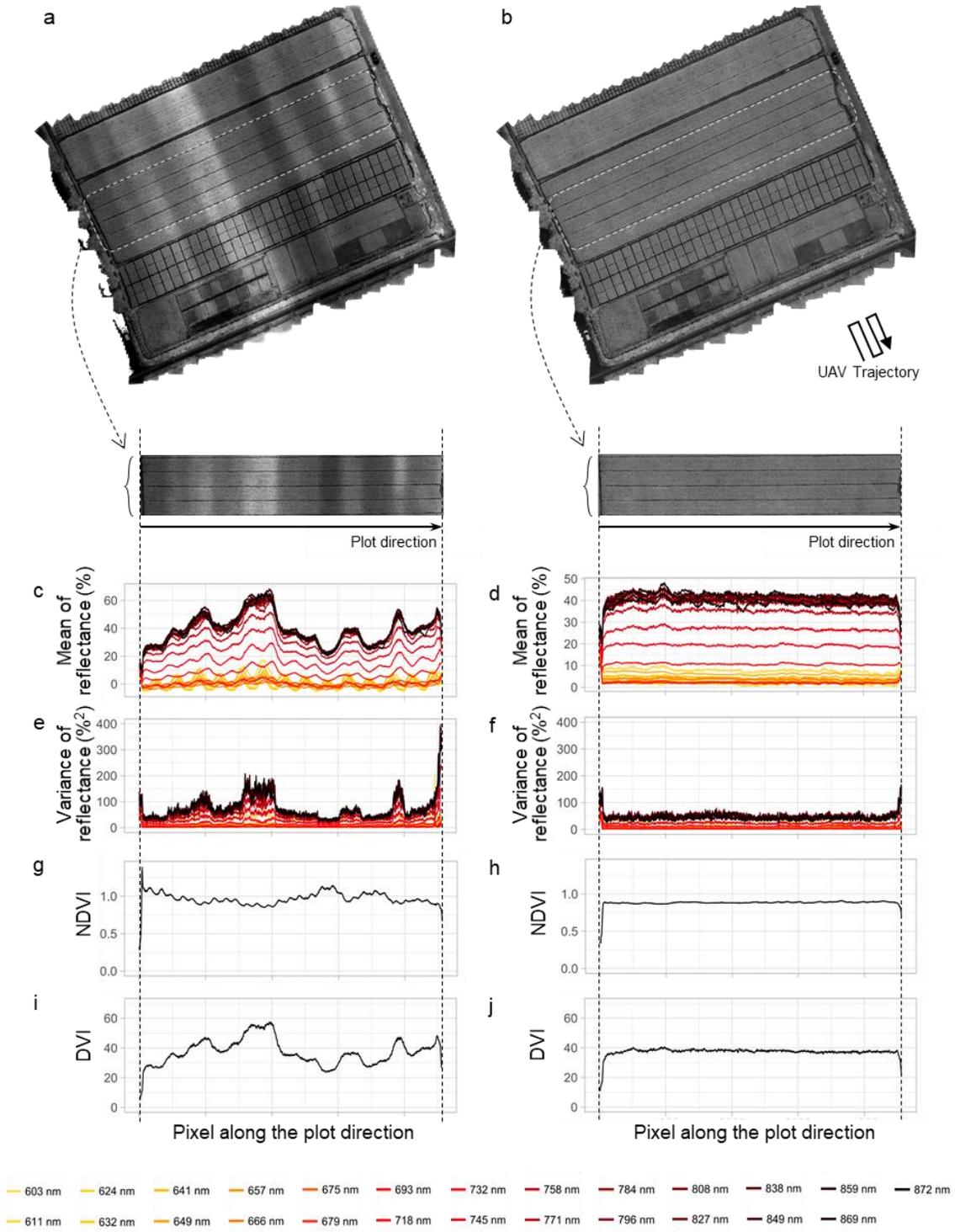

Fig. 10. Model validation in UAV an imaging campaign under fluctuating cloudy weather. Orthomosaic images of the crop field were constructed using images generated by (a) the empirical line

method (ELM) and (b) the 4-Band multiple linear regression (MLR) model. The arrow illustrates the UAV flight trajectory. White rectangles indicate the homogeneous crop plot for the follow-up examinations. Along the plot direction, the means and variances of reflectance values for all vertical pixels within the crop plot (as indicated by the brackets) are shown for the orthomosaic generated with (c, e) the ELM and (d, f) the 4-Band MLR model. Calculated from the vertical mean reflectance, the vegetation indices normalized difference vegetation index (NDVI) and difference vegetation index (DVI) are displayed for the orthomosaic generated with (g, i) the ELM and (h, j) the 4-Band MLR model. For illustration, the band at a wavelength of 827 nm from the multispectral imagery is displayed in greyscale.

Upon examining a long strip of a crop field with homogeneous vegetation and assessing the vertical means of reflectance along the plot direction, it was found that higher reflectance corresponded to higher pixel intensity in the ELM-corrected orthomosaic, and vice versa (Fig. 10c). When averaged across bands, the CV for the vertical means of reflectance along the plot direction was calculated to be 146.6%. However, with the implementation of the 4-Band MLR model, the reflectance spectra for each band were stabilized over the homogeneous vegetation area, resulting in a CV of 7.3%. This represents a substantial 95.0% improvement in consistency (Fig. 10d). Notably, in the ELM-corrected orthomosaic, negative reflectance values appeared in the short wavelength range due to the linear transformation applied to all pixels using temporally mismatched reference panel reflectance. Also, the low signal-to-noise ratio in these bands resulted in particularly large errors along the seam lines of the orthomosaic. These issues were not present in the orthomosaic constructed using the 4-Band MLR model. Furthermore, the vertical variance of reflectance exhibited abrupt changes along the seam lines in the ELM-corrected orthomosaic, but these variations were not present in the orthomosaic built with the 4-Band MLR model (Fig. 10e-f).

Finally, two vegetation indices, NDVI and DVI, were calculated using vertical means of reflectance values along the plot direction at wavelengths of 679 nm and 796 nm, representing the red and near-infrared bands, respectively. In the ELM-corrected orthomosaic, the CV for NDVI and DVI was 6.8% and 22.1%, respectively. In contrast, with the implementation of the 4-Band MLR model, the CV for NDVI and DVI was reduced to 1.0% and 2.1%, respectively, reflecting improvements in consistency of 86.0% and 90.3% (Fig. 10g-j). This demonstrates that reliable measures of reflectance can enhance the accuracy of vegetation indices, both in ratio and non-ratio forms.

## 4. Discussion

### 4.1 Importance of obtaining accurate reflectance from UAV spectral imagery

UAV remote sensing demonstrates considerable potential in large-scale agricultural applications. Consequently, a growing number of lightweight spectral sensors have been developed for use on UAVs, as they provide an efficient and straightforward means for data collection with minimal time and financial investment in planning (Maes and Steppe, 2019). Unlike indoor spectral measurements on benchtops, UAV remote sensing occurs in uncontrolled environments, where light interaction with plant canopies are significantly affected by factors such as weather, lighting conditions, and solar position (Fig. 3; Fig. 4; Supplementary Fig. S6). Without an effective method for generating or calibrating reflectance, data may suffer from substantial errors induced by environmental conditions, which can adversely impact downstream spectral analysis and potential applications of UAV spectral imaging in remote sensing communities. Alternatively, UAV remote sensing might need to be restricted to specific time windows or illumination conditions, reducing its convenience for task planning. This study addresses these challenges by stabilizing measurements of downwelling solar spectrum, thereby enabling the generation of high-quality reflectance from multispectral imagery collected over a broad sampling window.

Various methods have been proposed for extracting plant biochemical and physiological traits or agronomic traits from UAV spectral imagery, including machine learning algorithms, statistical models, and radiative transfer models. However, the transferability of these models across datasets collected under different environmental conditions remains a topic of debate. For example, an analysis of PROSPECT model inversion performance indicated that proportional error affected results as varying levels of random noise and bias were introduced to reflectance spectra in a synthetic dataset (Sun et al., 2018), which aligns with our own observations (data not shown). While enhancing model robustness involves multiple facets, providing accurate reflectance values is a crucial first step.

In this study, performance of the proposed 4-Band MLR model was comprehensively validated in the ground experiment under partially cloudy conditions (Fig. 8) and the UAV experiment validation at different times of the day (Fig. 9), resulting in RMSE values of 2.24% and 2.03%, respectively. In addition, implementation of the model improved the consistency in reflectance prediction by 95.0% during the validation of the UAV experiment under significant cloud movement dynamics (Fig. 10). It is worth noting that these validation experiments exhibit substantial variability in time, space and environmental conditions, in comparison to similar studies in the literature (Suomalainen et al., 2021; Xue et al., 2023). Thus, the effectiveness and robustness of the 4-Band MLR model were sufficiently demonstrated.

Additionally, an extensive body of literature has used UAV multispectral imagery to convert reflectance into vegetative indices, which serve as proxies to estimate phenotypic traits (Xie and Yang, 2020). This preference arises because these indices, often expressed as ratios, effectively mitigate the substantial noise introduced by varying environmental conditions during imaging (Bannari et al., 1995). Common examples include NDVI, enhanced vegetation index (EVI), green-normalized difference vegetation index (GNDVI), and normalized green-red difference index (NGRDI). Statistical surveys reveal that non-ratio vegetation indices appear in only 2% of the literature on vegetative indices related to RGB or multispectral imagery (Radočaj et al., 2023). In this study, the DVI, a non-ratio form vegetation index, was evaluated alongside the popular vegetation index, NDVI, in a UAV experiment under significant cloud movement dynamics. Results demonstrated a substantial improvement of 86.0% and 90.3% for these two indices, respectively, in terms of prediction consistency after implementing the 4-Band MLR model, suggesting that reliable estimations can be achieved for both ratio and non-ratio form vegetation indices (Fig. 10). Thus, with more accurate reflectance generation across diverse measurement environments, there is significant potential to explore the use of non-ratio vegetation indices, allowing for a more comprehensive extraction of agricultural information from spectral data.

## 4.2 Comparisons between different methods

As a widely adopted technique for radiometric correction, ELM is limited by its inability to track real-time reference data for calibration during UAV imaging campaigns, leading to instability under cloudy conditions (Aasen et al., 2018; Wang, 2021). Our evaluation confirmed this, showing ELM had a high RMSE exceeding 12% under fluctuating cloudy conditions (Fig. 8). ELM performed well in sunny conditions, with an RMSE of 1.21%, only slightly worse than the best-performing models. This is consistent with a previous research, where ELM achieved an RMSE of 3% to 5% in clear-sky and overcast conditions, respectively (Daniels et al., 2023). However, the method's strong performance in stable lighting environments does not compensate for its significant susceptibility to cloudy weather. In real-world scenarios, the probability of clear weather is very limited. Cloud fraction over land was estimated to be as high as 55%, according to a study based on 12 years of continuous observations from satellites (King et al., 2013). Typically, in Hangzhou, where this study was conducted, clear-sky days account for less than 10% of the summer season (Weather Spark), highlighting a major drawback for using ELM as a robust calibration method in UAV spectral sensing.

In our study, the DLS method was simulated by integrating over the full spectral range of the DS. The results indicated an RMSE of 3.44% under sunny conditions and 3.30% under fluctuating cloudy conditions (Fig. 8). Although the spectrometer was mounted on a gimbal to ensure stability, this advantage was not reflected in the RMSE compared to findings from another study evaluating DLS performance in clear and overcast skies, where RMSE ranges from 3% to 5% (Daniels et al., 2023). This discrepancy may be attributed to the broader environmental variations encountered during our multi-day diurnal evaluation, which included a range of sunny and fluctuating cloudy conditions. Yet, consistent with the referenced study, DLS performance showed minimal variation across different illumination conditions, suggesting that with stabilized reference light intensity, real-time corrections can be effective across varying weather scenarios. As expected, the error decreased with the inclusion of additional PCs in the PCR models, reducing the RMSE by 58.3% when three additional PCs were incorporated into the Top-1PC PCR model (Fig. 8). Due to the high inter-correlation among DN values across wavelengths, the first PC displayed uniform loadings throughout the spectrum (Fig. 7), acting as a broad measure of total light intensity, similar to the DLS method used in this study. Consequently, the Top-1PC PCR model yielded results nearly identical to those of the DLS method.

The direct correction method, Top-4PC PCR model, and 4-Band MLR model, despite being derived through different methodologies, exhibit commonalities in two key areas. First, their performance metrics for RMSE and rRMSE were closely aligned, with no significant differences detected from statistical analysis. Second, all three models demonstrated approximately 45% better performance in sunny weather compared to fluctuating cloudy conditions. Among these, the direct correction method is considered the most traditional yet effective method, as both the DS and MSI utilize narrow-band spectra. The comparable performance of the Top-4PC PCR model indicates that reflectance correction can be effectively achieved by capturing variations and modeling shared features embedded in the reference spectra. Additionally, the PCR model offers an advantage over direct correction method by allowing corrections of target bands outside the wavelength range of the reference spectra.

Surprisingly, the 4-Band MLR model achieved performance equivalent to that of the direct correction and Top-4PC PCR model, while relying on fewer hardware conditions. Specifically, the 4-Band MLR model utilizes only four selected bands as references for correcting all target bands, in contrast to the other methods which either use the same amount of matched bands in reference as the target bands, or the full spectrum from a spectrometer's output. Moreover, the spectral resolution of the reference spectra used in the 4-Band MLR model is significantly lower, with a bandwidth of 30 nm, compared to the 6 nm resolution employed by the other methods. Notably, the bands at 722 nm and 915 nm are sensitive to water vapor, and the band at 773 nm is sensitive to both water vapor and oxygen gas due to atmospheric absorption (Ortenberg, 2010). This suggests that the 4-Band MLR model effectively manages to minimize the redundancies present in reference spectra collected under varying weather conditions, and reflects significant patterns and features with just a few key bands. More importantly, this signal remains valid even when measured at a lower spectral resolution.

## 4.3 Key considerations in reflectance generation

The first key factor to consider is the spatio-temporal synchronization between the target object and the reference data. Ideally, incident light illumination should be captured in real-time and on-site. While "real-time" synchronization is relatively straightforward to achieve through time registration between devices, "on-site" synchronization is more challenging. In our study, although the downwelling optical sensor is mounted on the UAV and travels with it, a horizontal gap exists between the light beam hitting the vegetation target and the light intercepted by the on-board optical sensor. This gap equal to the UAV flight height multiplied by the arctangent of the solar altitude angle (Fig. 11). For a UAV flight height of 15 m and a minimum solar altitude angle of 40 degrees in our experiment, the horizontal gap is calculated to be 11.5 m. Theoretically, the error caused by this gap is minimized around noon and increases as the solar altitude angle decreases. Such a gap causes minimal error in incident light measurement on sunny days, where the lighting environment is stable over a large area, but it can be problematic under fluctuating cloudy conditions (Barker et al., 2020). The extent of the error depends on the thickness and movement speed of the clouds. Therefore, for UAV imaging campaigns conducted under fluctuating cloudy weather, it is advisable to plan the flights at higher solar altitude angles to minimize the measurement error. Further strategies are needed to optimize this approach.

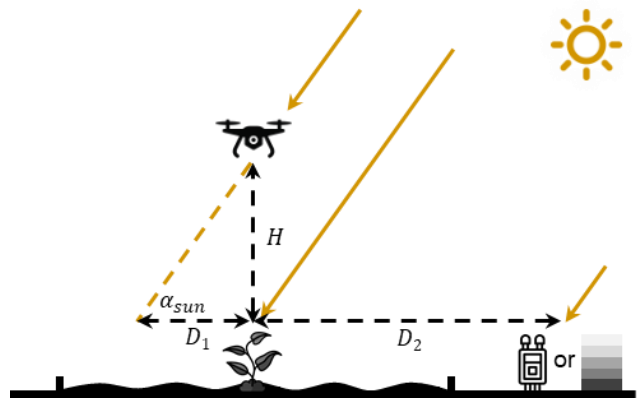

Fig. 11. Diagram illustrating the lighting conditions of UAV remote sensing in field environments. H: UAV flight height; $\alpha_{sun}$ : solar altitude angle; D1: horizontal gap between light beam hitting vegetation target and light beam intercepted by optical sensor mounted on UAV; D2: horizontal gap between light beam hitting vegetation target and light beam intercepted by optical sensor or reference panels placed on the ground.

Nonetheless, compared to the ELM method, which relies on reference data that is temporally disconnected from the imagery of targets, or methods that collect real-time reference data from locations potentially hundreds of meters away from the imaging target (Xue et al., 2023), the UAV's onboard light interception provides a substantial improvement. By positioning the reference sensor directly on the UAV, desynchronization between reference data and target measurements is significantly minimized (Daniels et al., 2023; DeCoffe et al., 2022). It is worth noting that attempts were also made to collect reference data at the exact location of the imaging target by positioning reference panels throughout the field and imaging one panel at short time intervals (Nigon et al., 2020). While this approach can be effective for relatively small UAV imaging campaigns, it can quickly become time-consuming and costly for setup when scaled to larger areas. It should also be noted that, at a flight height of 15 m, the multispectral imager used in our study covers a ground area of 11.0 m in width. This results in a heterogeneous lighting environment within a single image under fluctuating cloudy conditions. Addressing anisotropic reflectance correction at the pixel level remains a challenging problem. If such errors are acceptable, a solar altitude angle of 40 degrees can serve as a general guideline for determining the operational window of our method, as the horizontal gap of 11.5 m at this angle closely approximates the imaging area of 11.0 m. These conditions translate to an approximately 7-hour operational window in the middle of the day for UAV flights during summer at a latitude of 30° N.

The second key factor for ensuring accuracy in reflectance generation when using a secondary optical sensor as a reference is the stability and precision of downwelling sunlight acquisition. Researchers have noted that if the optical sensor is mounted directly on a UAV, variations in UAV attitude during flight can lead to significant errors in light measurement due to changes in viewing angles (Adão et al., 2017; Taddia et al., 2020). The previous study has reported significant errors in reflectance values adjusted by DLS, which were associated with varying wavelengths and the reflectance of reference panels (Deng et al., 2018). Additionally, errors exceeding 10% were observed for blue and near-infrared bands on cloudy and partially cloudy days when using the DLS correction (Barker et al., 2020). However, satisfactory results can be achieved if the DLS tilting is corrected or if the DLS is fixed on the ground (Burkart et al., 2014; Suomalainen et al., 2021). It was reported that during a UAV imaging campaign, pitch and roll angles can deviate by up to 15° and 10°, respectively, and the UAV's attitude may not accurately reflect the angle information for the mounted optical sensor (Olsson et al., 2021). Thus, stabilizing the downwelling sensor, or providing corrections for its readings based on changes in light angles is critical. In this study, the DS was mounted on an upward gimbal with a wind shield to prevent air disturbances, and additional correction on the cosine corrector was performed to minimize errors from interactions between the corrector unevenness and solar positions. These measures are essential for ensuring precise and consistent incident light measurements.

The third key factor in calibrating or generating reflectance on UAV platform is the cost. Mini-sized spectrometers can cost around $1,000. However, our results indicate that the reference spectra were more than sufficient in terms of band number and spectral resolution. The 4-Band MLR model, using just four selected bands (722 nm, 773 nm, 800 nm, and 915 nm) with a 30 nm bandwidth, performed as well as the more expensive direct correction model and Top-4PC PCR model. This suggests that a costly spectrometer can be replaced with four optical filtered light detectors, significantly reducing hardware expenses. A fair-quality light detector and a custom optical filter cost approximately $10 and $20 each, respectively, totaling around $120 for four sensors. Adding the $30 cost for the gimbal and $100 cost for microcomputer, the total expense for reflectance correction with 4-Band MLR model is about $250, making it a cost-effective option for broader implementation.

The fourth key factor is convenience of application. ELM requires placing reference panels close to the imaging area, which can be cumbersome due to their size and weight. Finding suitable locations for these panels can be challenging, especially in terrains like forests or hills, and panels can become dirty over time. In contrast, an onboard optical sensor is compact and integrates easily with the UAV. In our setup, the gimbal, spectrometer, and microcomputer were powered by the UAV, requiring no additional equipment. This approach simplifies operation, minimizes labor, and allows the entire system to be transported conveniently in the same suitcase as the UAV.

## 4.4 Limitations and future works

This study focused on generating reflectance from DN values for multispectral images, but extracting precise reflectance for the target requires additional corrections. Bidirectional reflectance distribution function (BRDF) models, which account for incident light angle, reflection angle, canopy surface properties, and other environmental factors, are necessary for accurate reflectance correction within the scene (Shell, 2004). To avoid overcomplicating the issue, our study assumed Lambertian characteristics for all reference panels and crop canopies. However, this assumption may be violated (Burkart et al., 2015). Previous studies suggest field targets exhibit Lambertian properties when the solar altitude angle is between 45 to 60 degrees and imaging is at nadir (DeCoffe et al., 2022). Although extensive studies have been conducted on related topics, a standardized correction processing pipeline is still lacking and requires further research (Aasen et al., 2018).

In our study, a snapshot MSI with 25 spectral bands was used, but our reflectance generation models can theoretically be applied to line-scan (pushbroom) hyperspectral imagers, which are also common on UAV platforms. The core approach of aligning imager spectra with on-board reference spectra is adaptable, though data formats may vary. While the 4-Band MLR model provided good results even beyond its coverage range (Table 2), its performance across a broader wavelength range (e.g., 400 nm to 2500 nm) still requires further evaluation. Additional factors, such as $CO_2$ gas absorption around 2000 nm, may also need to be considered (Ortenberg, 2010).

Additionally, this study was limited to UAV remote sensing within low-altitude airspace, assuming negligible atmospheric absorption between the UAV altitude of 15 m and the ground. Previous research has demonstrated significant atmospheric absorption across a range of altitudes, from below 20 m to as high as 150 m, and has addressed this issue using direct correction method (Suomalainen et al., 2021). Although atmospheric absorption correction at varying flight altitudes was not explored in this study, it would be valuable to investigate whether our proposed method can be extended to account for such effects.

Optimizations in hardware can be pursued in several areas. Firstly, our current setup triggers the DS and MSI independently. Synchronizing the triggers for both devices is recommended for improved coordination and data efficiency. Secondly, exposure times were fixed before takeoff, which may cause overexposure if lighting conditions change rapidly during flight. Implementing automatic exposure adjustments could mitigate this issue and optimize signal-to-noise ratios. Thirdly, although we achieved a gimbal return-to-center time of about 6 seconds at each UAV turn, gimbal instability can still have minor effect over short distances in the subsequent flight line. To address this, we expanded the imaging area on each edge in our flight planning. Further research is needed to tackle the factors affecting gimbal sensitivity.

## 5. Conclusions

This study revealed that a significant amount of variation is associated with environmental factors such as weather, lighting conditions, and solar position. The relative change in solar intensity reached 214.0% in a span of 30 min on a fluctuating cloudy day, highlighting the necessity of real-time tracking of incident light as correction references for reflectance generation. By integrating a DS on top of the UAV along with the MSI, this study allows for spatiotemporal synchronization between data collected from the two sensors. Unlike previous reports, where the downwelling optical sensor was fixed directly on the UAV, we mounted it on an upward gimbal equipped with a wind shield to mitigate air disturbances, thereby improving the measurement stability of the DS. Additionally, correction of the DS measurements was undertaken to eliminate errors introduced by solar positions. These carefully implemented measures are crucial for obtaining precise and consistent incident light measurements, serving as an essential foundation for reference data in reflectance generation. PCA on the solar spectrum acquired from a broad range of environmental conditions indicated that most variations in the solar spectrum are embedded in wavelength ranges with strong atmospheric absorption of water vapor and oxygen gas. As a result, PCR models were constructed using major PCs, and an MLR model was developed, selecting 4 bands from the sensitive wavelength range (722 nm, 773 nm, 800 nm, 915 nm). Comparing the proposed models with conventional ones, we found that the Top-4PC PCR model and the 4-Band MLR model performed equally well with the direct correction model. Using RMSE as an evaluation metric, they outperformed the DLS method by approximately 59% across environmental conditions and outperformed the ELM by approximately 86% under fluctuating cloudy weather, demonstrating a significant advantage in model accuracy. Among the three, the 4-Band MLR model was preferred due to its lower hardware requirements. It can operate with four light detectors equipped with optical filters with a bandwidth of 30 nm, significantly reducing the cost of implementation. This work also represents a notable case where the performance of the proposed method, the 4-Band MLR model, is evaluated over significant temporal and spatial scales with dynamically changing illumination conditions. Ground validation under partial cloud cover resulted in RMSE and rRMSE of 2.24% and 7.2%, respectively. UAV validations conducted across various time points in sunny weather resulted in averaged RMSE and rRMSE of 2.03% and 6.6%. Moreover, the consistency of reflectance measured in a large homogeneous rice crop field under significantly fluctuating cloud movements improved by 95.0% with the implementation of the 4-Band MLR model. In summary, this study presents a robust, reliable, and cost-effective approach for reflectance generation, facilitating standardized data collection across various illumination conditions and significantly extending the operational time window for UAV multispectral remote sensing missions.

## Author contribution

**Jiayang Xie**: Software, Methodology, Formal analysis, Writing - Original Draft, Writing - Review & Editing, Visualization. **Yutao Shen**: Software, Methodology, Validation, Writing - Review & Editing. **Haiyan Cen**: Writing - Review & Editing, Supervision, Project administration.

## Declaration of competing interest

Declaration of Competing Interest The authors declare that they have no competing financial interests or personal relationships that could have appeared to influence the work reported in this paper.

## Data availability

Data will be made available on request.

## Acknowledgement

This work was funded by the National Key R & D Program of China (2021YFD2000104), National Natural Science Foundation of China (32371985), and Fundamental Research Funds for the Central

Universities (226-2022-00217). We would like to thank Mengqi Lyu, Xiaoyue Du and Xuqi Lu for their support with data collection and Dr. Haowei Fu, Dr. Youfa Li and Dr. Gui Chen for providing the field experimental site at Jiaxing Academy of Agricultural Sciences.# References

Aasen, H., Burkart, A., Bolten, A., Bareth, G., 2015. Generating 3D hyperspectral information with lightweight UAV snapshot cameras for vegetation monitoring: From camera calibration to quality assurance. ISPRS Journal of Photogrammetry and Remote Sensing 108, 245–259. https://doi.org/10.1016/j.isprsjprs.2015.08.002

Aasen, H., Honkavaara, E., Lucieer, A., Zarco-Tejada, P.J., 2018. Quantitative Remote Sensing at Ultra-High Resolution with UAV Spectroscopy: A Review of Sensor Technology, Measurement Procedures, and Data Correction Workflows. Remote Sensing 10, 1091. https://doi.org/10.3390/rs10071091

Adão, T., Hruška, J., Pádua, L., Bessa, J., Peres, E., Morais, R., Sousa, J., 2017. Hyperspectral Imaging: A Review on UAV-Based Sensors, Data Processing and Applications for Agriculture and Forestry. Remote Sensing 9, 1110. https://doi.org/10.3390/rs9111110

Arroyo-Mora, J.P., Kalacska, M., Løke, T., Schläpfer, D., Coops, N.C., Lucanus, O., Leblanc, G., 2021. Assessing the impact of illumination on UAV pushbroom hyperspectral imagery collected under various cloud cover conditions. Remote Sensing of Environment 258, 112396. https://doi.org/10.1016/j.rse.2021.112396

Ballester, C., Hornbuckle, J., Brinkhoff, J., Smith, J., Quayle, W., 2017. Assessment of In-Season Cotton Nitrogen Status and Lint Yield Prediction from Unmanned Aerial System Imagery. Remote Sensing 9, 1149. https://doi.org/10.3390/rs9111149

Bannari, A., Morin, D., Bonn, F., Huete, A.R., 1995. A review of vegetation indices. Remote Sensing Reviews 13, 95–120. https://doi.org/10.1080/02757259509532298

Barker, J.B., Woldt, W.E., Wardlow, B.D., Neale, C.M.U., Maguire, M.S., Leavitt, B.C., Heeren, D.M., 2020. Calibration of a common shortwave multispectral camera system for quantitative agricultural applications. Precision Agric 21, 922–935. https://doi.org/10.1007/s11119-019-09701-6

Baugh, W.M., Groeneveld, D.P., 2008. Empirical proof of the empirical line. International Journal of Remote Sensing 29, 665–672. https://doi.org/10.1080/01431160701352162

Beisl, U., Braunecker, B., Cramer, M., Driescher, H., Eckardt, A., Fricker, P., Gruber, M., Hilbert, S., Jacobsen, K., Jagschitz, W., Jahn, H., Kirchhofer, W., Leberl, F., Neumann, K., Sandau, R., Schoenermark, M., Tempelmann, U., 2010. Digital Airborne Camera. Introduction and Technology.

Bergmüller, K.O., Vanderwel, M.C., 2022. Predicting Tree Mortality Using Spectral Indices Derived from Multispectral UAV Imagery. Remote Sensing 14, 2195. https://doi.org/10.3390/rs14092195

Burkart, A., Aasen, H., Alonso, L., Menz, G., Bareth, G., Rascher, U., 2015. Angular Dependency of Hyperspectral Measurements over Wheat Characterized by a Novel UAV Based Goniometer. Remote Sensing 7, 725–746. https://doi.org/10.3390/rs70100725

Burkart, A., Cogliati, S., Schickling, A., Rascher, U., 2014. A Novel UAV-Based Ultra-Light Weight Spectrometer for Field Spectroscopy. IEEE Sensors Journal 14, 62–67. https://doi.org/10.1109/JSEN.2013.2279720

Caturegli, L., Corniglia, M., Gaetani, M., Grossi, N., Magni, S., Migliazzi, M., Angelini, L., Mazzoncini, M., Silvestri, N., Fontanelli, M., Raffaelli, M., Peruzzi, A., Volterrani, M., 2016. Unmanned Aerial Vehicle to Estimate Nitrogen Status of Turfgrasses. PLoS ONE 11, e0158268. https://doi.org/10.1371/journal.pone.0158268

Cui, M., Sun, Y., Huang, C., Li, M., 2022. Water Turbidity Retrieval Based on UAV Hyperspectral Remote Sensing. Water 14, 128. https://doi.org/10.3390/w14010128

Daniels, L., Eeckhout, E., Wieme, J., Dejaegher, Y., Audenaert, K., Maes, W.H., 2023. Identifying38